\documentclass[aps,pre,superscriptaddress]{revtex4} 
\usepackage[utf8]{inputenc}
\usepackage[T1]{fontenc}
\usepackage{lmodern}
\usepackage{graphicx}
\usepackage{dcolumn}
\usepackage{bm}
\usepackage{color}
\usepackage{multirow}

\begin{document}

\title{Preferential Concentration of Inertial Sub-Kolmogorov Particles. \\
		\large{The roles of mass loading of particles, $St$ and $Re_{\lambda}$ numbers}}
       

\author{Sholpan Sumbekova}
\email{sholpan.sumbekova@legi.grenoble-inp.fr}
\affiliation{Universit?? Grenoble Alpes, LEGI, F-38000 Grenoble, France} 
 \affiliation {CNRS, LEGI, F-38000 Grenoble, France}
 
 \author{Alain Cartellier}%
 \email{alain.cartellier@legi.grenoble-inp.fr}
\affiliation{Universit?? Grenoble Alpes, LEGI, F-38000 Grenoble, France} 
\affiliation {CNRS, LEGI, F-38000 Grenoble, France}

\author{Alberto Aliseda}%
 \email{aaliseda@u.washington.edu}
\affiliation{Department of Mechanical Engineering, University of Washington, Seattle WA 98195-2600, USA}

\author{Mickael Bourgoin}
\email{mickael.bourgoin@ens-lyon.fr}
\affiliation{Univ Lyon, Ens de Lyon, Univ Claude Bernard, CNRS, 
Laboratoire de Physique, F-69342 Lyon, France}



\date{April 1st, 2016}

\begin{abstract}
Turbulent flows laden with inertial particles present multiple open questions and are a subject of great interest in current research. Due to their higher density compared to the carrier fluid, inertial particles tend to form high concentration regions, i.e. clusters, and low concentration regions, i.e. voids, due to the interaction with the turbulence. In this work, we present an experimental investigation of the clustering phenomenon of heavy sub-Kolmogorov particles in homogeneous isotropic turbulent flows. Three control parameters have been varied over significant ranges: $Re_{\lambda} \in [170 - 450]$, $St\in [0.1 - 5]$ and volume fraction $\phi_v\in [2\times 10^{-6} - 2\times 10^{-5}]$. The scaling of clustering characteristics, such as the distribution of Vorono\"i areas and the dimensions of cluster and void regions, with the three parameters are discussed. In particular, for the polydispersed size distributions considered here, clustering is found to be enhanced strongly (quasi-linearly) by $Re_{\lambda}$ and noticeably (with a square-root dependency) with $\phi_v$, while the cluster and void sizes, scaled with the Kolmogorov lengthscale $\eta$, are driven primarily by $Re_{\lambda}$. Cluster length $\sqrt{\langle A_c \rangle}$ scales up to $\approx 100 {\eta}$, measured at the highest $Re_{\lambda}$, while void length $\sqrt{\langle A_v \rangle}$  scaled also with  $\eta$ is typically two times larger ($\approx 200 {\eta}$). The lack of sensitivity of the above characteristics to the Stokes number lends support to the "sweep-stick" particle accumulation scenario. The non-negligible influence of the volume fraction, however, is not considered by that model and can be connected with collective effects.

\end{abstract}

\maketitle
\section{Introduction}
Turbulent flows laden with inertial particles can be found in a broad range of engineering systems and geophysical phenomena. Droplets in clouds, cleaning sprays, aerosol pollutants, marine snow and planetesimals are just a few examples of such flows. Unlike tracer particles, inertial particles do not follow the flow velocity, but rather have their own dynamics resulting from the complex interaction of the particle inertia, their gravitational settling velocity and the fluid excitation across the continuous turbulent spectrum. The behaviour of turbulent flows laden with inertial particles is an active field of theoretical, numerical and experimental research. An important and unique aspect of inertial particles interacting with a turbulent background is their tendency to cluster, creating a inhomogeneous particle concentration field, a phenomenon known as \emph{preferential concentration}. Atlhough several mechanisms, such as the centrifugal expulsion of particles from the core of the eddies~\cite{bib:maxey1987_JFM} or the sticking property of zero-acceleration points of the carrier flow~\cite{bib:goto2008_PRL,bib:coleman2009_PoF}, have been proposed to explain preferential concentration, no clear picture has emerged yet regarding the scaling and dominant parameters controlling the underlying physical processes at play. This lack of quantitative understanding considerably limits our capacity to build physical models to describe and predict the phenomenon and its consequences,  for instance coalescence or evaporation/condensation of droplets, in practical situations. Empirical models are also difficult to develop as preferential concentration involves many ingredients whose specific roles have not been clearly identified yet: particle inertia, turbulence characteristics, gravitational settling, disperse phase volume fraction, etc.. In this context, the present article reports a systematic experimental exploration of preferential concentration as several control parameters known to influence this phenomenon are varied over a wide range. These physical parameters can be related to several dimensionless parameters:
\begin{itemize}
\item Inertia is characterized by the particle Stokes number $St=\tau_p/\tau_\eta$, the ratio between the particle viscous relaxation time $\tau_p=\frac{1}{18}\frac{\rho_p}{\rho_f}\frac{{d_p}^2}{\nu}$ (where $d_p$ is the drop diameter, $\rho_p$ and $\rho_f$ denote the particle and carrier fluid densities respectively, and $\nu$ is the fluid's kinematic viscosity) and the dissipation time of the carrier turbulence $\tau_\eta=\sqrt{\nu/\epsilon}$ (where $\epsilon$ is the turbulent kinetic energy dissipation rate). For small particles (with diameter $d_p$ much smaller than the dissipative scale of the flow $\eta=\left(\nu^3/\epsilon\right)^{1/4}$, the Stokes number can be related to the particle to fluid density ratio $\Gamma=\rho_p/\rho_f$ and to the ratio $\Phi$ between the particle diameter $d_p$ and the Kolmogorov microscale $\eta$ as $St=\frac{\Phi^2}{36}(1+2\Gamma)$. The Stokes number is therefore bounded in this analysis by the assumption of small particles ($d/\eta <<1$). Given a density ratio value (800 for water droplets in air) the Stokes number can not be larger than 5 (for a maximum value of $d/\eta \approx 0.3$). Values of Stokes number beyond this value represent large particles that can not be studied based only on the Stokes number, but require an indepedent measure of their finite size, as shown by the impact of finite-sized neutrally-buoyant particles on the turbulent characteristics of the carrier flow~\cite{bib:qureshi2007_PRL,bib:qureshi2008_EPJB, bib:xu2008_PhysicaD,bib:lucci2011_PoF,bib:homann2010_JFM,bib:fiabane2012_PRE,bib:doychev2010_ICFM}.
. 

\item The strength of the turbulent excitation on the particles is related here to the Reynolds number of the carrier flow $Re=\sigma_u L /\nu$ (with $\sigma_u$ and $L$ equal to the velocity rms and the correlation length of the velocity fluctuations, respectively). In the present work, we use the Reynolds number based on the Taylor micro-scale $R_\lambda=\sqrt{15 Re}$. 

\item Gravitational settling characterized by the non-dimensional ratio of the particle terminal velocity (taken as Stokes settling velocity in still fluid for this $Re_p<1$ particles) to the eddy velocity scale of the vortices that interact most strongly with the particles, in this case the Kolmogorov velocity~\cite{bib:wang1993_JFM,bib:yang1998_JFM,bib:aliseda2002_JFM}. This parameter represents the influence of crossing trajectories effects~\cite{bib:csanady1963_JAtmSc,bib:Wells83} and preferential sweeping effects~\cite{bib:maxey1987_JFM}) on the interaction of inertial particles with the carrier turbulence. These interactions impact the settling rate of the particles as well as their clustering properties. 

\item The overall concentration of the disperse phase in the flow is characterized by the volume fraction $\phi_v$ occupied by the particles. Volume fraction is known to impact particle-turbulence interactions at various levels. In dilute situations, $\phi_v \leq 10^{-5}$, it is primarily the turbulence that affects the particles dynamics with no global modification of the properties of the turbulent carrier flow due to the presence of the particles, the \emph{one-way coupling}  regime. At higher volume fractions, $10^{-5}<\phi_v<10^{-3}$, \emph{two-way coupling} effects emerge with a modification of the carrier turbulence due to the presence of the particles. At even higher disperse phase concentrations, $10^{-3}<\phi_v$,  \emph{four-way} coupling mechanisms with additional particle-particle interactions appear. The volume fraction values used in the experiments are well within the dilute, one way coupling regime $\phi_v \leq 2 \times ~10^{-5}$ and the mass loading is always less than 2 percent, so that no significant modification of the turbulence in the carrier phase is considered.
\end{itemize}

\begin{figure}[t]
\centering
\includegraphics[width=0.75\columnwidth]{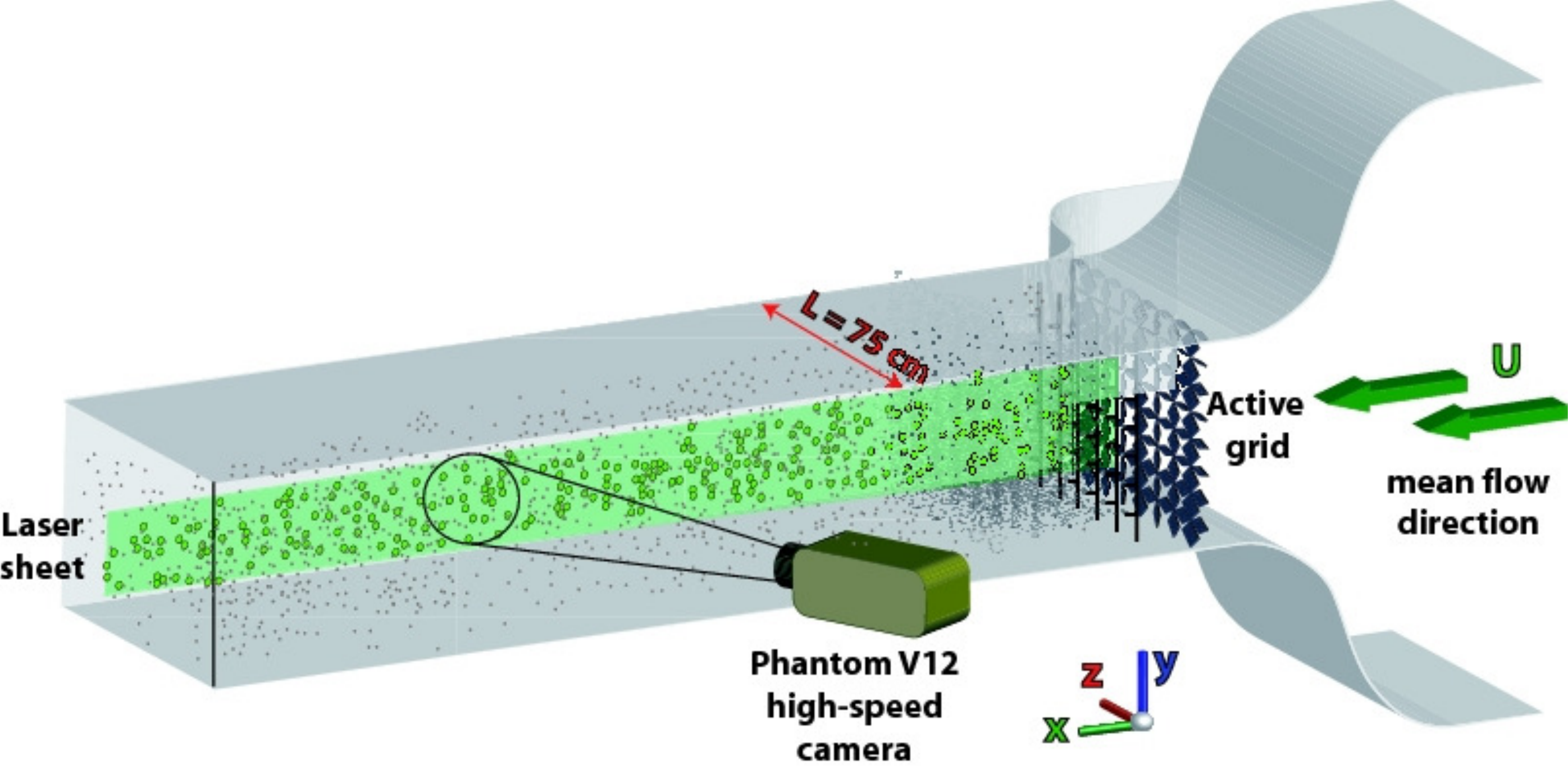}
\caption{Schematic view of the experimental facility.}
\label{fig:experiment}
\end{figure}

Other parameters, such as polydispersity (both in particle size and/or density), particle's shape anisotropy, etc. can also influence clustering properties, but will not be addressed here.

As previously mentioned, although clear evidence of clustering modification with the control parameters  studied here has been shown in experiments and simulations, a quantitative measure of the impact of each of these parameters on preferential concentration over a wide range of values has not been obtained to date. For instance, available numerical studies (mostly carried under the assumption of point particles~\cite{bib:maxey1983}) and the few available experiments, indicate that the Stokes number directly influences the clustering phenomenon, with a maximum degree of clustering for particles with $St={\cal{O}}(1)$.  Existing results also suggest that clustering level increases with increasing Reynolds number of the carrier flow~\cite{bib:obligado2014_JoT}. Similarly, it was recently shown that the disperse phase volume fraction has a non-trivial effect on clustering~\cite{bib:monchaux2010_PoF}, with a non-linear dependency of the accumulation within clusters with the global concentration (even in situations of one-way coupling, where no global modulation of the carrier turbulence is expected due to the presence of the particles). Aliseda \emph{et al.}~\cite{bib:aliseda2002_JFM} have also shown that gravitational settling is non-trivially connected to the preferential concentration phenomenon and to the global volume fraction and can be collectively enhanced within clusters. A better insight into such behaviors is required in order to clearly disentangle the role of Stokes number, Reynolds number and volume fraction and eventually start paving the way towards possible strategies to develop predictive and accurate models of preferential concentration. 

One of the difficulties to characterize the specific role of these parameters clearly, lies in the practical complexity to unambiguously and systematically disentangle their specific contribution in actual experiments. For instance, for a given class of particles (fixed size and density), varying the Reynolds number of the carrier flow (for instance by reducing the viscosity $\nu$ of the fluid or increasing the energy dissipation rate $\epsilon$) also results in a change of the particle Stokes number (as the dissipation scale $\eta=(\nu^3/\epsilon)^{1/4}$, and hence the ratio $\Phi=d/\eta$ also varies). Similarly, regarding the volume fraction $\phi_v$, even if it is varied within the one-way coupling regime ($\phi_v\leq 10^{-5}$), the number density within clusters may be larger, due to preferential concentration, inducing subtler particle/turbulence and particle/particle interactions, which may in turn result in non-trivial collective dynamics of particles and of clusters of particles out of reach for the commonly-used point particle models.




\begin{figure}[t]
\centering
\includegraphics[scale=0.4]{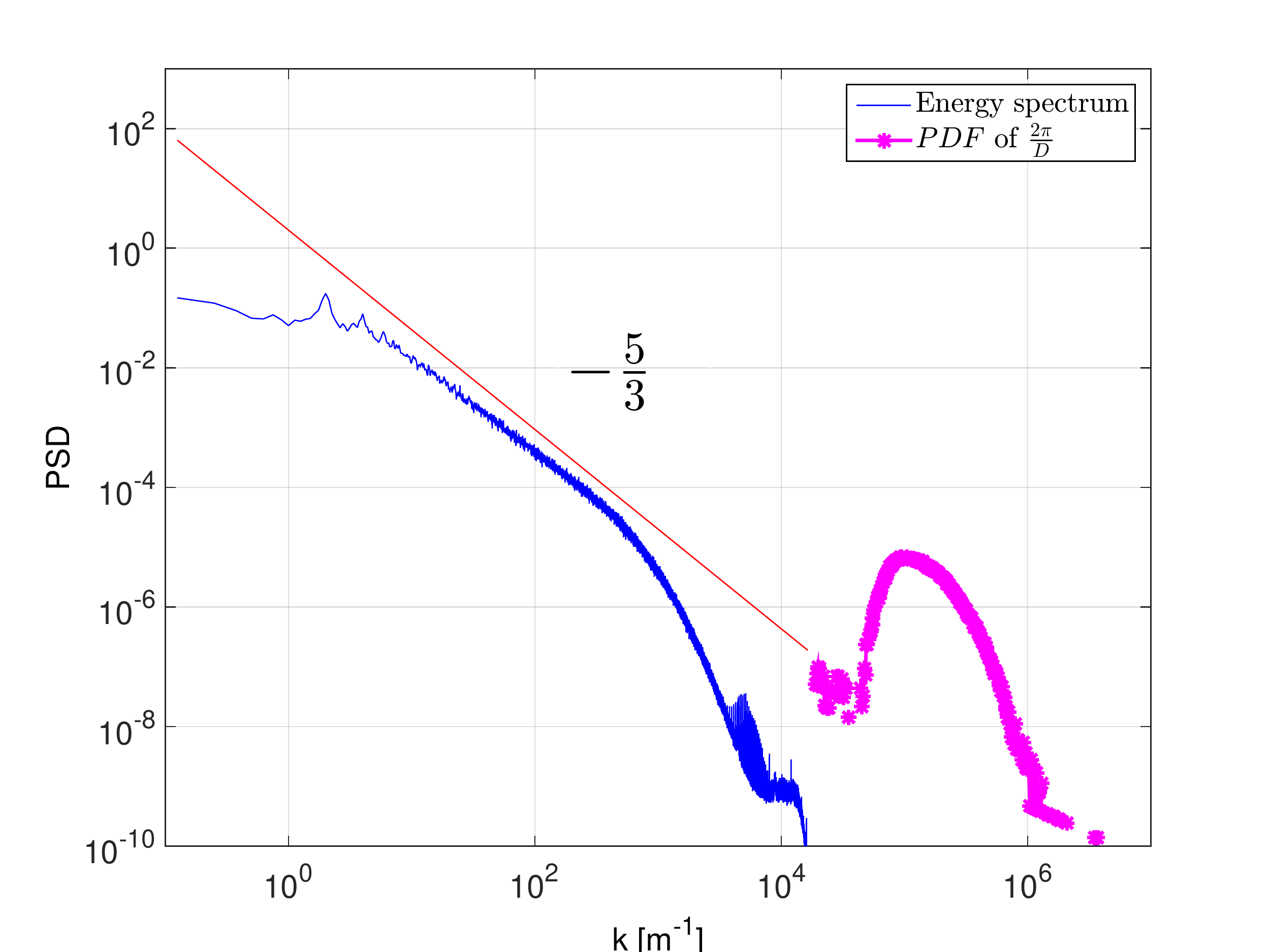}
\caption{Typical Eulerian energy spectrum of velocity fluctuations produced downstream of the active grid, obtained from classical hot-wire anemometry, at a distance of 3.5~m downstream the grid, corresponding to the position where water droplets preferential concentration is investigated. {The typical distribution of $2\pi/D$, where $D$ is the droplets diameter, shown in magenta on this graph (with arbitrary units in the ordinate axis) demonstrates that all droplets are indeed smaller than the Kolmogorov length scale.}}
\label{fig:spectrum}
\end{figure}

We present an experimental investigation of preferential concentration of water droplets in homogeneous isotropic turbulence turbulence generated by an active-grid, where $St$, $R_\lambda$ and $\phi_v$ have been varied independently (within a wide range if experimental available values). We report the dependency of the degree of preferential concentration based on Vorono\"i tesselation analysis~\cite{bib:monchaux2010_PoF,bib:monchaux2012_IJMF}, and of cluster and void geometry. The article is organized as follows: section~\ref{sec:exp} presents the experimental facility and the strategy to explore the $(St, R_\lambda,\phi_v)$ parameter space; section~\ref{sec:voro} details the Vorono\"i tesselation method and proposes new strategies to better handle possible bias of this analysis (due to illumination inhomogeneity in the experiment, for example); in section~\ref{sec:results}, we report the main results of this investigation, before proposing a detailed discussion in section~\ref{sec:discussion}, with conclusions and future lines of research identified from this work.
\section{Experimental setup}\label{sec:exp}
\indent Experiments were conducted in a wind tunnel with a test section of 0.75~m$\times$0.75~m$\times$4~m (see fig.~\ref{fig:experiment}). Homogeneous isotropic turbulence is produced with an active grid located at the entrance of the test section. The mean streamwise velocity $U$ in the wind tunnel was varied in the range $U\in [2.5 - 10]$~m/s (corresponding turbulence properties are given below in Table~\ref{tab:turb}). Water droplets are injected 15~cm downstream of the active grid using an array of 18 pressure injection nozzles supplied with a controlled flow rate of water via a high pressure pump. Three injector sizes (with different orifice diameters $D_{inj} = 0.3,~0.4,~0.5~mm$) were used in order to vary the size distribution of droplets injected in the flow. The droplet volume fraction can be further controlled by varying the flow rate of water $F_{water}$ injected, which in our experiment evolves in the range $F_{water}\in[0.8-1.9]$~L/min. Overall, the combination of the three control parameters $(U,D_{inj},F_{water})$ allows us to explore the parameter space $(St,R_\lambda,\phi_v)$. The  main properties of the carrier turbulence, of the seeded water droplets and the accessible parameter space is described in the following subsection.

\begin{table*}[t]
\centering
\caption{Conditions of the experimental runs.}
\label{tab:turb}
\begin{tabular}{cccccccccccccc}
\hline
\\
$D$ & $F_{water}$ & $U$ & $u\prime$ & $\eta$ & $\tau_{\eta}$ &  $\epsilon$ & $D_{max}$ & $std(D)$ & $D_{32}$ & $St_{D_{max}}$ & $St_{max}$& $\phi_v$ & $Re_{\lambda}$  \\
$[mm]$ &$[l/min]$& $[m/s]$ & $[m/s]$ & $[\mu m]$ & $[s]$ & $[m^2/s^3]$ & $[\mu m]$ &$[\mu m]$ & $[\mu m]$ &  & &  $\times 10^{-4}$ &  \\
\\
\hline
\\
\multirow{8}{*}{0.3} & \multirow{4}{*}{0.8} & 2.36 & 0.30 & 431  &0.011  & 0.20  & 35                  & 17     & 60  & 0.3 & 0.26 & 0.09                     & 175                   \\
                     &                      & 4.11 & 0.59 & 250  & 0.004  & 1.35  & 32                  & 19     & 61  & 0.7 & 0.5 & 0.05                     & 259                    \\
                     &                      & 6.42 & 1.00 & 162  &   0.002 &  6.12 & 24                  & 18     & 60  & 1.0 & 0.29 & 0.03                     & 356                   \\
                     &                      & 9.19 & 1.49 & 114    &0.001 &20.73 & 37                  & 18     & 58  & 5.0 & 2.3 & 0.02                     & 459                  \\
                     \\
                     & \multirow{4}{*}{1.2} & 2.37 & 0.33 & 429 & 0.011 & 0.21   & 21                  & 17     & 52  & 0.1 & 0.05 & 0.14                     & 175                  \\
                     &                      & 4.03 & 0.62 & 255   &0.004 & 1.26  & 27                  & 17     & 59  & 0.5 & 0.17 & 0.08                     & 256                    \\
                     &                      & 5.95 & 0.92 & 174    & 0.002 & 4.74 & 24                  & 17     & 57  & 0.9 & 0.22 & 0.06                     & 337                    \\
                     &                      & 8.85 & 1.41 & 118 &0.001 & 18.27     & 28                  & 16     & 57  & 2.7 & 2.15 & 0.04                     & 447                   \\ 
                     \\
                     \hline
                     \\
\multirow{7}{*}{0.4} & \multirow{4}{*}{1.9} & 2.32 & 0.32 & 439   & 0.012& 0.19  & 36                  & 19     & 62  & 0.3 & 0.2 & 0.22                     & 172                    \\
                     &                      & 3.95 & 0.56 & 260   & 0.004& 1.17  & 45                  & 20     & 65  & 1.4 & 0.68 & 0.13                     & 252                    \\
                     &                      & 5.95 & 0.95 & 174   & 0.002& 4.74  & 32                  & 20     & 66  & 1.6 & 0.42 & 0.09                     & 337                    \\
                     &                      & 8.64 & 1.40 & 121    & 0.001 & 16.83 & 30                  & 20     & 69  & 2.8 & 0.77 & 0.06                     & 439                   \\
                     \\
                     & \multirow{3}{*}{1.4} & 4.07 & 0.59 & 252   & 0.004&1.31  & 26                  & 21     & 66  & 0.5 & 0.19 & 0.10                     & 258                    \\
                     &                      & 6.27 & 1.00 & 166   & 0.002& 5.64 & 28                  & 21     & 67  & 1.4 & 0.94 & 0.06                     & 350                    \\
                     &                      & 8.82 & 1.43 & 118   & 0.001& 18.02 & 29                  & 20     & 66  & 2.7 & 1.84 & 0.04                     & 446                    \\
                     \\
                   \hline
                   \\
\multirow{4}{*}{0.5} & \multirow{4}{*}{1.9} & 2.30 & 0.30 & 442  & 0.012& 0.19  & 23                  & 23     & 67  & 0.1 & 0.05 & 0.23                     & 172                   \\
                     &                      & 4.13 & 0.58 & 249  & 0.004& 1.37   & 32                  & 23     & 72  & 0.8 & 0.4 & 0.13                     & 260                    \\
                     &                      & 6.16 & 0.96 & 168  & 0.002 & 5.32  & 37                  & 22     & 70  & 2.3 & 1.02 & 0.08                     & 345                    \\
                     &                      & 8.68 & 1.40 & 120  & 0.001 & 17.05  & 33                  & 21     & 70  & 3.5 & 2.3 & 0.06                     & 440           \\
                     \\
                     \hline
                     \\
\end{tabular}
\end{table*}

\subsection{Turbulence generation}

The details of the active grid in the wind tunnel have been published in~\cite{bib:obligado2014_JoT}. Briefly, it is made up of eight vertical and eight horizontal shafts on which square wings are mounted. Each axis is controlled individually by a stepper motor, so that the solidity of the grid can be actively and dynamically changed. This turbulence generation technique was first introduced by Makita~\cite{bib:makita1991} and has been reproduced in multiple studies in the literature (\cite{bib:mydlarski1996_JFM,bib:poorte2002_JFM}, among others). When a random forcing protocol is used to drive the rotation of the shafts and flapping of the wings, it generates stronger turbulence than a passive grid (turbulence intensity in our active grid flow is of the order of 15-20\%, while it is typically 2-4\% in passive grid wind tunnel turbulence) while keeping good homogeneity and isotropy. Figure~\ref{fig:spectrum} shows a typical spectrum (measured with classical hot-wire anemometry) of the  carrier flow velocity fluctuations for $U=10~m/s$, where a well defined inertial range can be clearly identified over about 2 decades in wavenumber space. Table~\ref{tab:turb} summarizes the main properties of the turbulence and the disperse phase, for the different values of mean stream velocity $U$.


\subsection{Water droplets generation}

The 18 spray nozzles for water injection are fixed on 8 vertical bars at the same positions in the cross section as the vertical bars of the active grid, in order to minimize the flow disturbance due to the injector array. Hot-wire anemometry shows that the presence of the injector array does not modify the turbulence properties at the measurement location (3.5~m downstream of the grid); the turbulence spectra with and without injectors are undistinguishable. 
The size distribution of the droplets is controlled by the injector orifice size  ($D_{inj} = 0.3,~0.4,~0.5~mm$) and depends only weakly on the flow rate $F_{water}$. Thus, the volume fraction $\phi_v$ was varied with a combination of four different liquid flow rates and four different wind tunnel speeds, for a range of $2~10^{-6}-2~10^{-5}$, without significantly affecting the droplet mean size, which was varied independently. The ranges of dimensional and non-dimensional parameter spanned in the experiments are shown in figure~\ref{fig:experimental_map}.

\begin{figure}[t]
\centering
\includegraphics[width=.75\columnwidth]{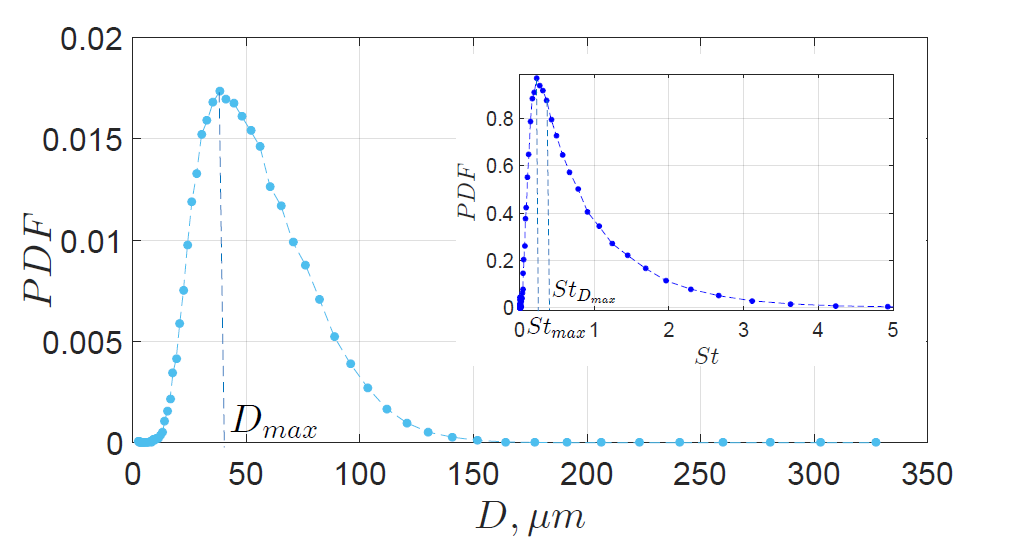}
\caption{Typical diameter distribution of water droplets in the wind tunnel (produced by the $0.4$~mm injectors at a flow rate of $1.9$~l/min). Inset: corresponding distribution of Stokes numbers for $R_\lambda = 172$. Dashed lines indicate the Stokes number defined on the most probable diameter and the most probable Stokes number.}
\label{fig:dropdist}
\end{figure}

\paragraph{Droplets Stokes number.} Fig.~\ref{fig:dropdist} shows a typical size distribution of water droplets obtained from Phase Doppler Interferometry (PDI) measurements. The spray is strongly polydispersed, with a well-defined most probable diameter. The droplet size distributions have been measured using PDI for all the considered experimental conditions. The most probable diameter $D_{max}$, the Sauter mean diameter $D_{32}$ and the standard deviation $std(D)$ are reported in Table~\ref{tab:turb}. The most probable diameter evolves in a narrow range, from $21$ to $45~\mu m$. Otherwise, all conditions exhibit a comparable degree of polydispersity as $std(D)/D_{32} = 0.31 \pm 10 \%$ and $std(D)/D_{max} = 0.66 +50\% -30\%$. The inset in figure~\ref{fig:dropdist} represents the distribution of particle Stokes numbers corresponding to the size distribution shown.  
Values of Stokes span the range from $0.004$ to $20.7$, but the range $0.1$ to $2.01$ is consistently in every experiment reported here. A reference Stokes number for each experimental condition is defined using the most probable droplet diameter $D_{max}$ in the distribution. Thus, for each experimental condition, the most representative particle Stokes number is estimated as $St_{D_{max}}=\frac{\Phi_{max}}{36}(1+2\Gamma)$, with $\Phi_{max}=(D_{max}/\eta)^{2}$ and $\Gamma=\rho_{water}/\rho_{air}\simeq 830$. Alternative choices are possible. For example, one may also refer to the most probable Stokes number, $St_{max}$ (see inset in figure~\ref{fig:dropdist}) that slightly differs from the Stokes number based on $D_{max}$. For the experimental conditions explored here, the difference between the most probable $St_{max}$ and $St_{D_{max}}$  based on $D_{max}$ varies between $12 - 75 \%$ (see table~?\ref{tab:turb}). As pointed out earlier, since $St$ depends both on $D$ and $\eta$, for a given particle distribution the Stokes number varies with the flow Reynolds number. Thus, $St$ is sensitive to both wind tunnel speed and injector orifice size as experimental controls. 

\paragraph{Droplets volume fraction.} The volume fraction $\phi_v$ of the droplet disperse phase in the wind-tunnel is given by the ratio between the water flow rate $F_{water}$ through injector array and the total flow rate of air and water across the tunnel cross-section $F_{tot}=F_{air}+F_{water}$ with $F_{air}=S\cdot U$  (where $S=(0.75~m)^2\simeq 0.56$~m$^2$ is the area of the tunnel cross section): $\phi_v=\frac{F_{water}}{F_{tot}}$. Note that in all experiments $F_{water}< 2$~L/min while $F_{air} > 1.4\mathrm{ m^3/s} \gg F_{water}$, so that $F_{tot}\simeq F_{air}$ and $\phi_v\simeq\frac{F_{water}}{F_{air}}=\frac{F_{water}}{S\cdot U}$. Therefore, the volume fraction $\phi_v$ depends on both liquid  injection flow rate and the wind-tunnel mean speed.\\

\subsection{Parameter space}

\begin{figure*}[t]
\centering
\includegraphics[width=.49\columnwidth]{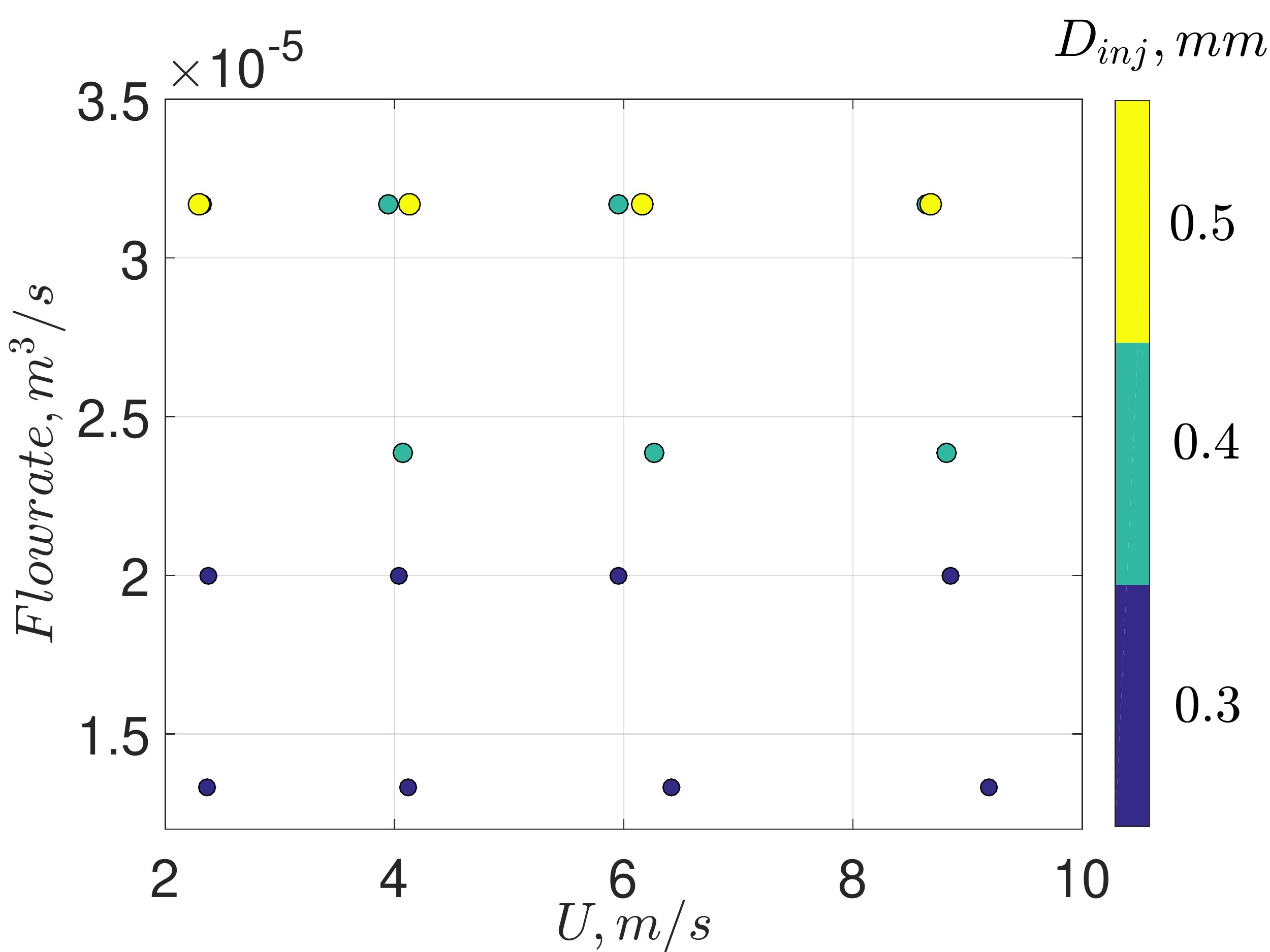}
\includegraphics[width=.49\columnwidth]{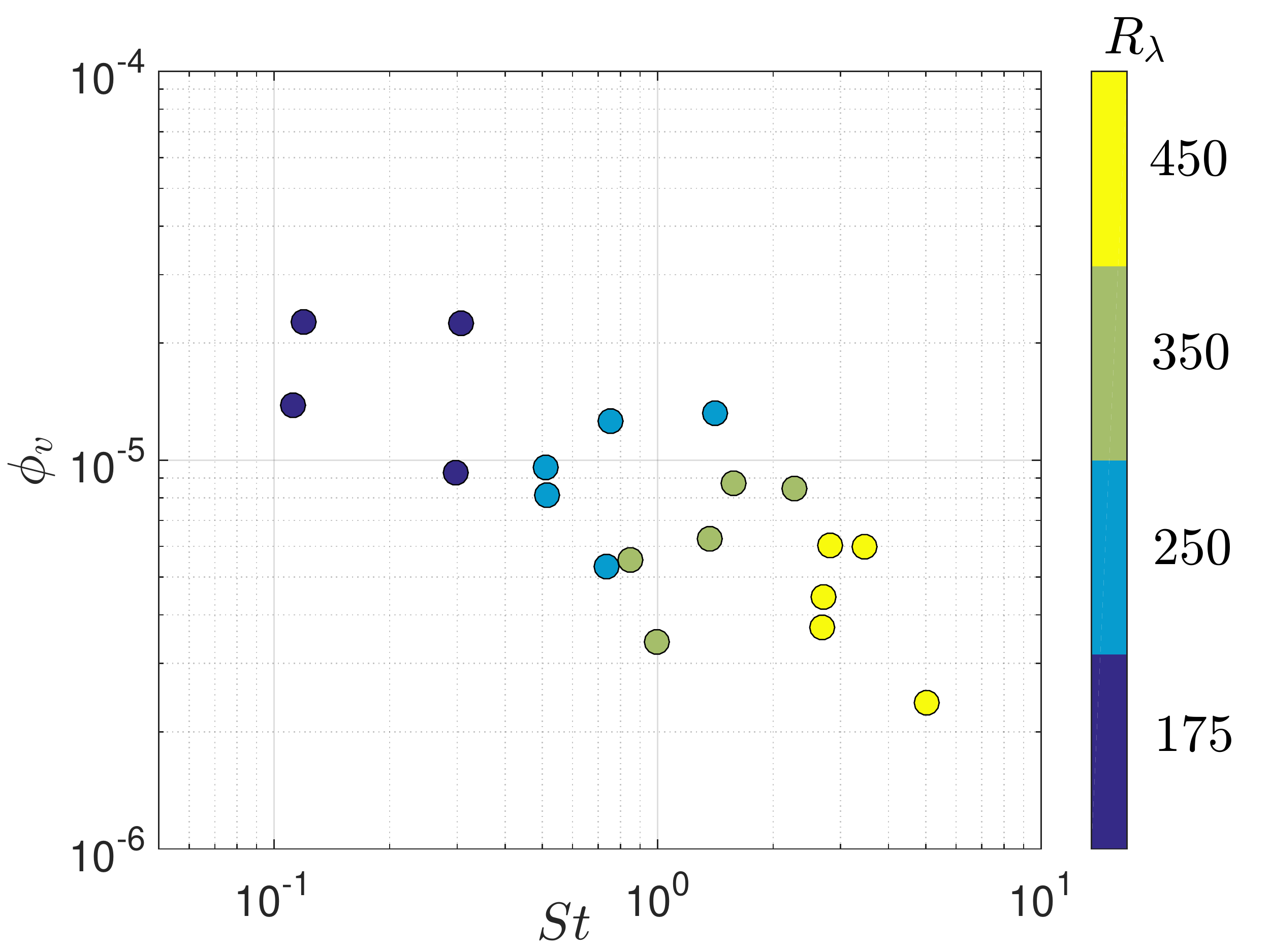} \\
(a) \hspace{.49\columnwidth} (b) 
\caption{(a) Map of explored control parameters. (b) Corresponding map of experimental parameter space.}
\label{fig:experimental_map}
\end{figure*}

The previous discussion shows the difficulty of independently varying the three parameters studied in this work: $(St,R_\lambda,\phi_v)$, as they are sensitive to changing more than one experimental control parameter $(U,D_{inj},F_{water})$. By independently varying the injector diameter, the liquid phase flow rate and the mean wind tunnel speed, the present study explores the parameter space $(St,R_\lambda,\phi_v)$ (see fig.~\ref{fig:experimental_map}) in the range: $St \in [0.1,5]$, $R_\lambda \in [170, 460]$ and $\phi_v \in [2 \times 10^{-6} - 2 \times 10^{-5}]$. Figures \ref{fig:experimental_map}a \& b represent the experimental parameters and the non-dimensional control parameters which were accessible in the experiments. 


However, in spite of these trends, there are still interesting portions and sections of the parameter space where 

By taking advantage of the experimental parameters that modified primarily one non-dimensional number with no or weak influence on the other two, the influence of one parameter could be investigated, tracing horizontal or vertical lines in fig.~\ref{fig:experimental_map}b. Thus, the effect of the Stokes number (horizontal lines) can be studied keeping a relatively constant volume fraction and with a moderate variation of Reynolds number. Similarly, the effect of volume fraction (vertical lines) can be teased out while keeping the Stokes and Reynolds numbers to small variations. Two sets of experimental conditions yielded almost identical Stokes number and volume fraction while changing the  Reynolds number by 50\%, hence allowing a limited exploration of the effect of Reynolds number.

\subsection{Data acquisition protocol}
The particles in the flow are illuminated by a laser sheet along the streamwise-vertical directions, at the midplane of the test section (fig.~\ref{fig:experiment}). The thickness of the laser sheet is $\approx1~mm$, or a few $\eta$. Because of the gaussian profile of the laser beam, the illumination is inhomogeneous in the vertical direction (a slight inhomogeneity also exists in the horizontal direction, mostly due to sheet formation near the waist of the laser). Therefore, the laser intensity is maximum at mid-height. Sequences of images are recorded using a high-speed camera (Phantom V12, Vision Research Inc., Wayne, N.J.). A 105~mm Nikon macro lens on a Scheimpflug mount was used to visualize the laser sheet in forward scattering conditions, improving the brightness of the droplets while keeping good focusing conditions over the entire image. The dimensions of the visualization area are $\simeq10~cm \times \simeq7~cm$ (covering a significant fraction of the integral scale of the carrier turbulence, which is of the order of $L_{int}\approx 15~cm$). For each experimental condition, defined by one triplet $(St,R_\lambda,\phi_v)$ in the parameter space in fig.~\ref{fig:experimental_map}, we record 20 movies at full resolution (1280x800 pixels) at an acquisition rate of 2600 frames per second; the duration of each movie is $\approx 3.27~s$ (8500 frames). The high frame rate was selected to enable particle tracking between consecutive images. The number of particles per image ranges typically between 500 and 2000 depending on flow conditions. \\


The measurement location is 3.5~m downstream the injection of the droplets, where turbulence is fully developed and sufficiently far away from the injection location for cluster formation to have reached a stationary state. The time required for clusters to form is indeed not well understood. If we consider the duration of transients observed in direct numerical simulations as an estimate, two options are available: either transients are of the order of the integral time scale $T_{int}$ of the carrier turbulence (provided that the particle response time is much smaller than $T_{int}$~\cite{bib:wang1993_JFM} as is the case in this study), or they depend on a combination of turbulent and particle characteristic time scales. For instance, Yang \& Lei~\cite{bib:yang1998_JFM} proposed 8 times the dissipation time scale $\tau_\eta$ plus 5 times the particle viscous realization time scale $\tau_p$. In all our experimental conditions, the transit time of droplets between the injection plane and the measurement location ranges from 60 to 600 particle response times, or several integral time scales (from $1.8$ to $2.5$). Thus, particle residence  time in the turbulence are expected  to be long enough for clusters to be have reached equilibrium.

\section{Vorono\"i tesselation analysis}\label{sec:voro}
\begin{figure}[t]
\includegraphics[width=.32\columnwidth]{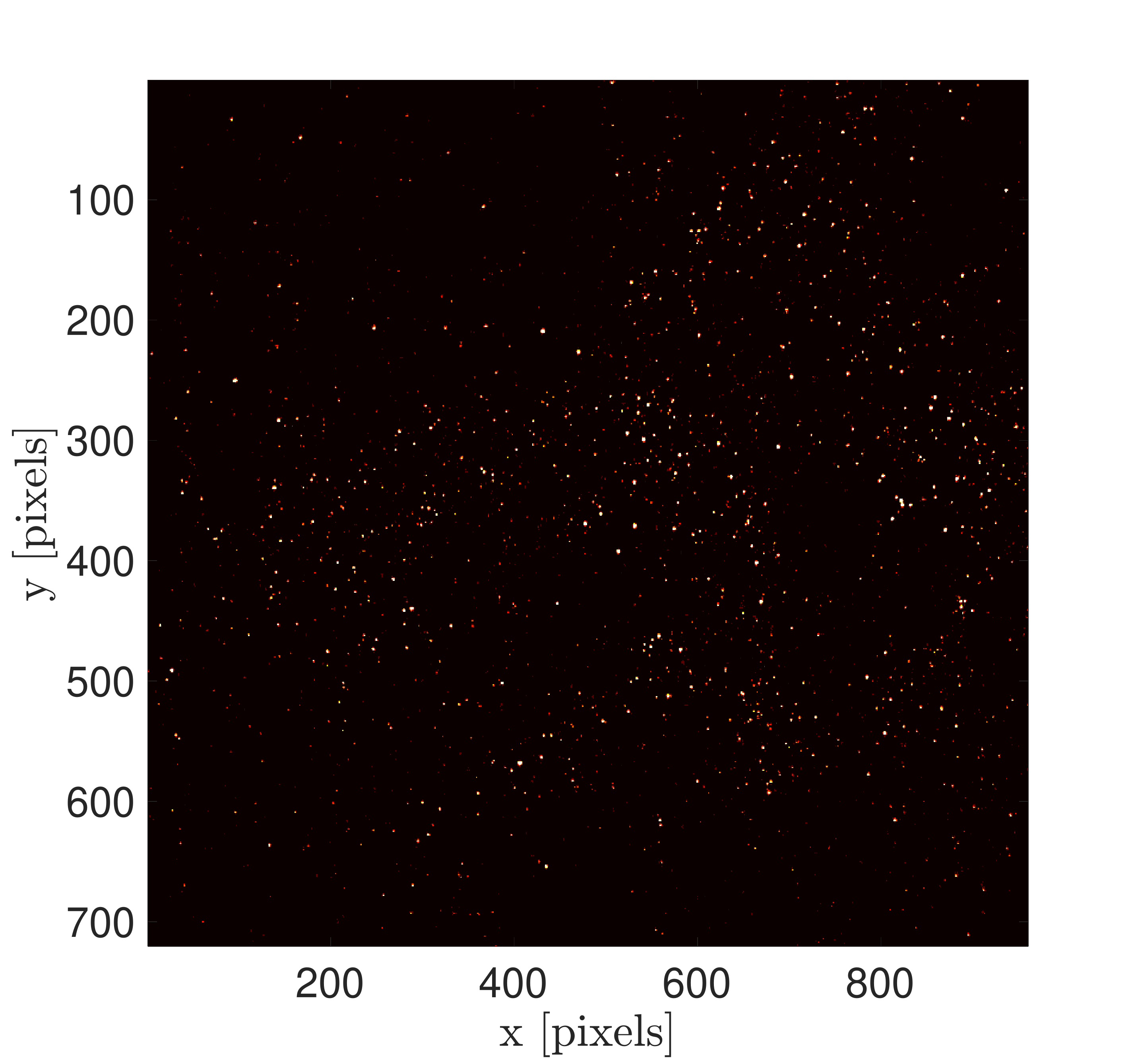}
\includegraphics[width=.32\columnwidth]{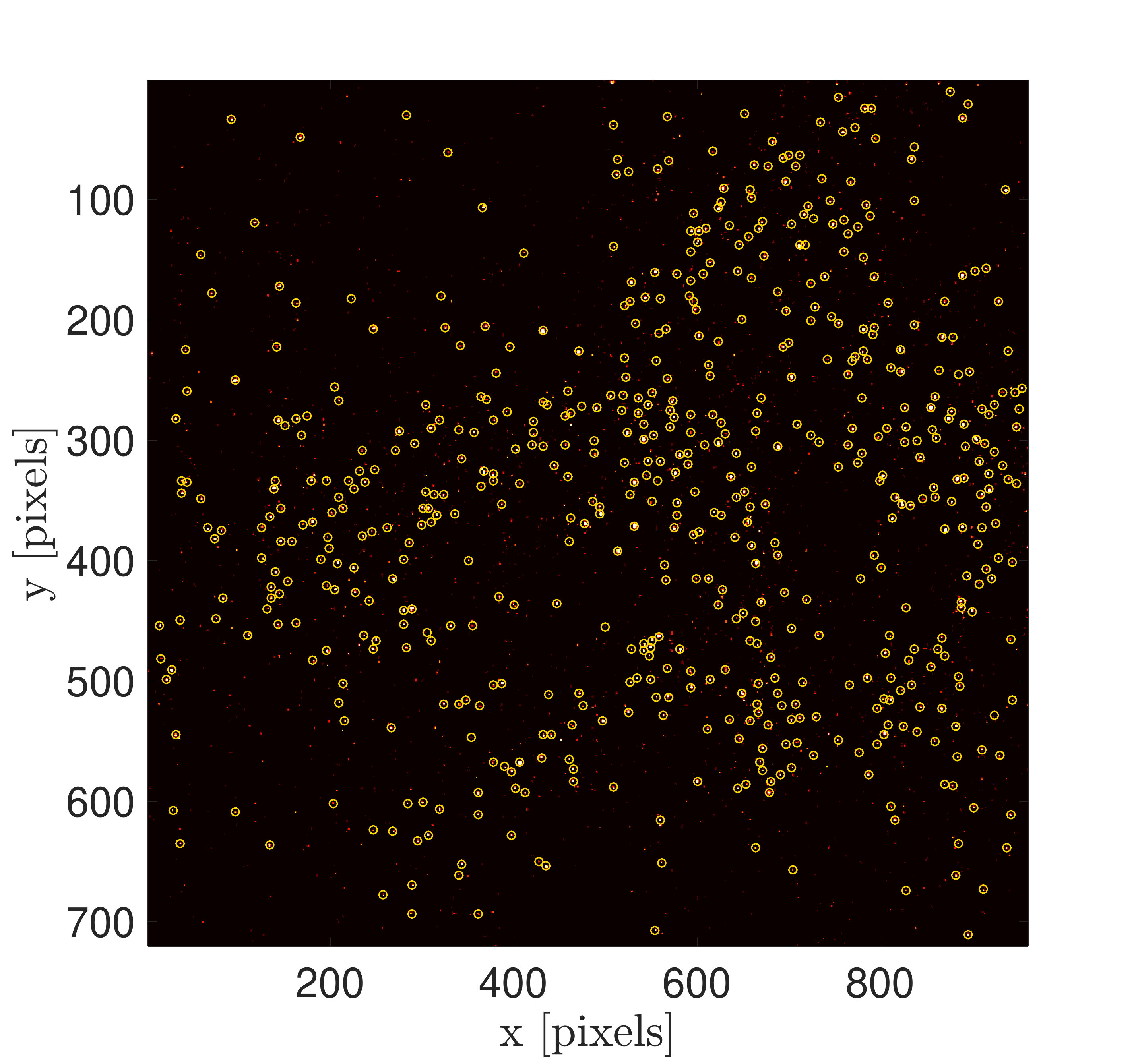}
\includegraphics[width=.32\columnwidth]{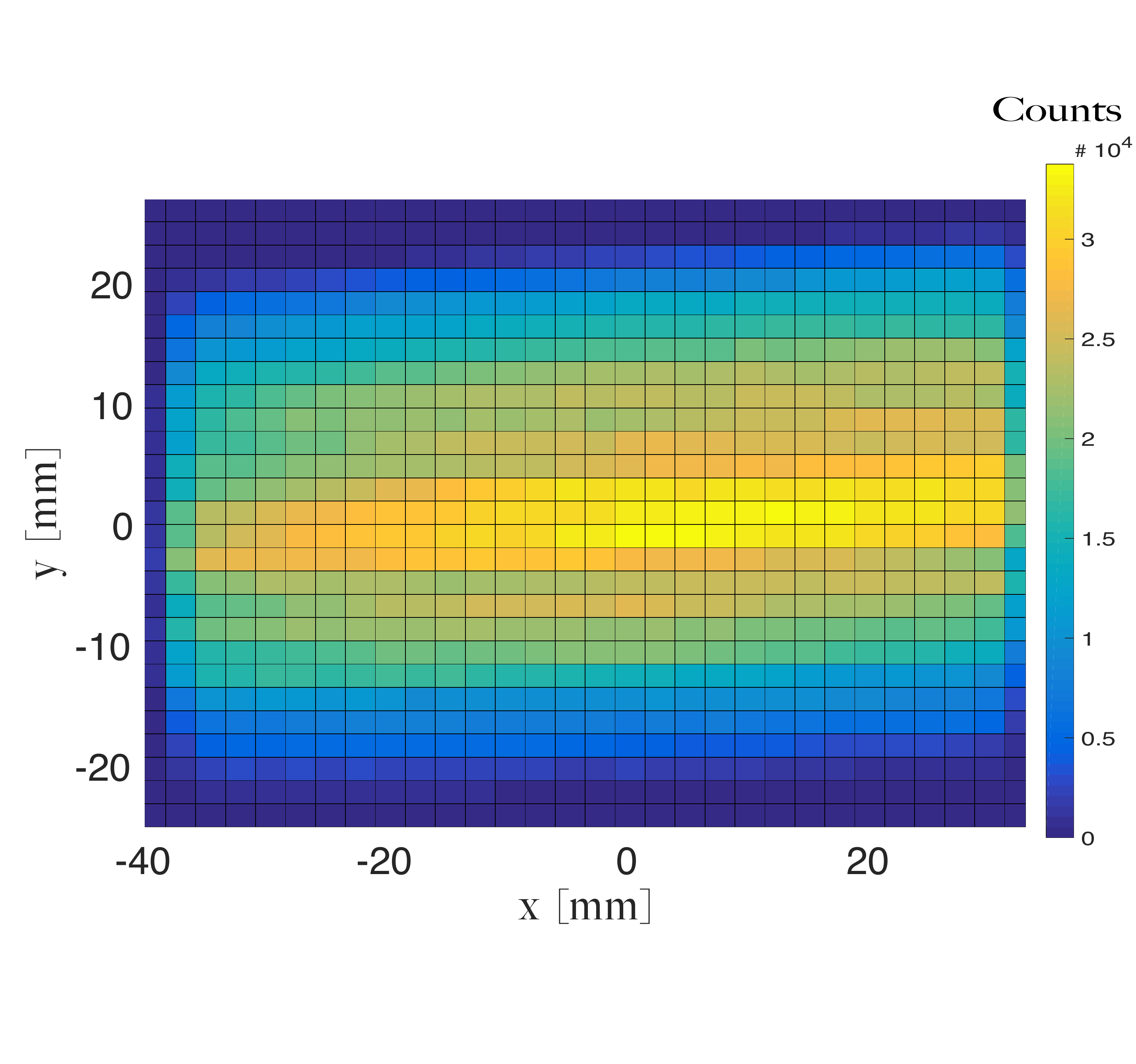}\\
(a) \hspace{.33\columnwidth} (b) \hspace{.33\columnwidth} (c)
\caption{(a) Typical raw image, (b) detected particles, (c) probability of detected particles. 
The large scale inhomogeneity of the detection reflects the gaussian intensity profile in the laser illumination.}\label{fig:raw_image}
\end{figure}

Vorono\"i tesselations, which have been proven to be a good estimator to quantify the clustering of particles~\cite{bib:ferenc2007_PhysA,bib:monchaux2010_PoF,bib:monchaux2012_IJMF,bib:tagawa2012_JFM,bib:uhlmann2014_JFM,bib:obligado2014_JoT} is used to diagnose the appearance and the importance of preferential concentration.

\begin{figure*}[t]
\centering
      \includegraphics[width=.32\columnwidth]{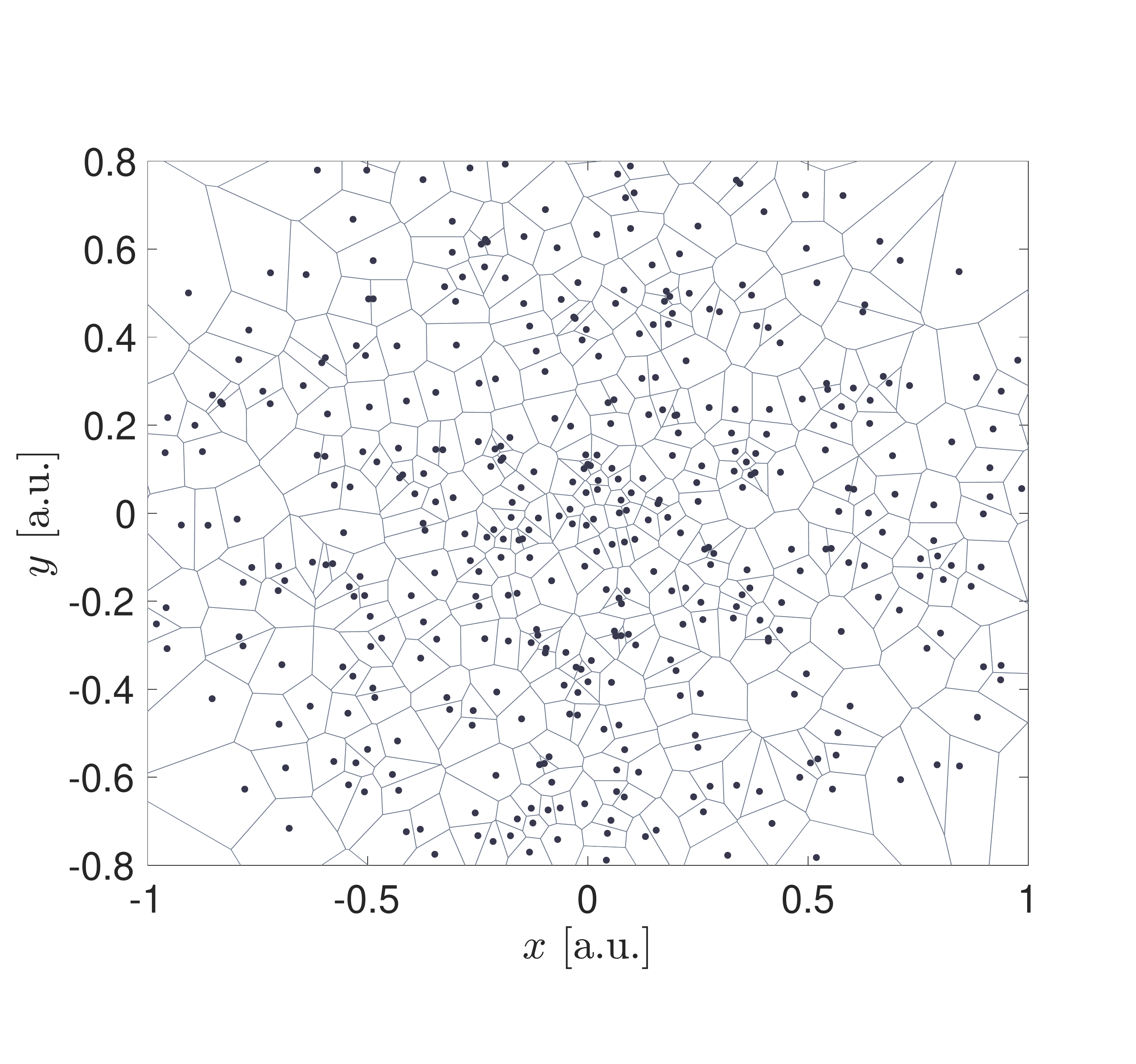} 
      \includegraphics[width=.32\columnwidth]{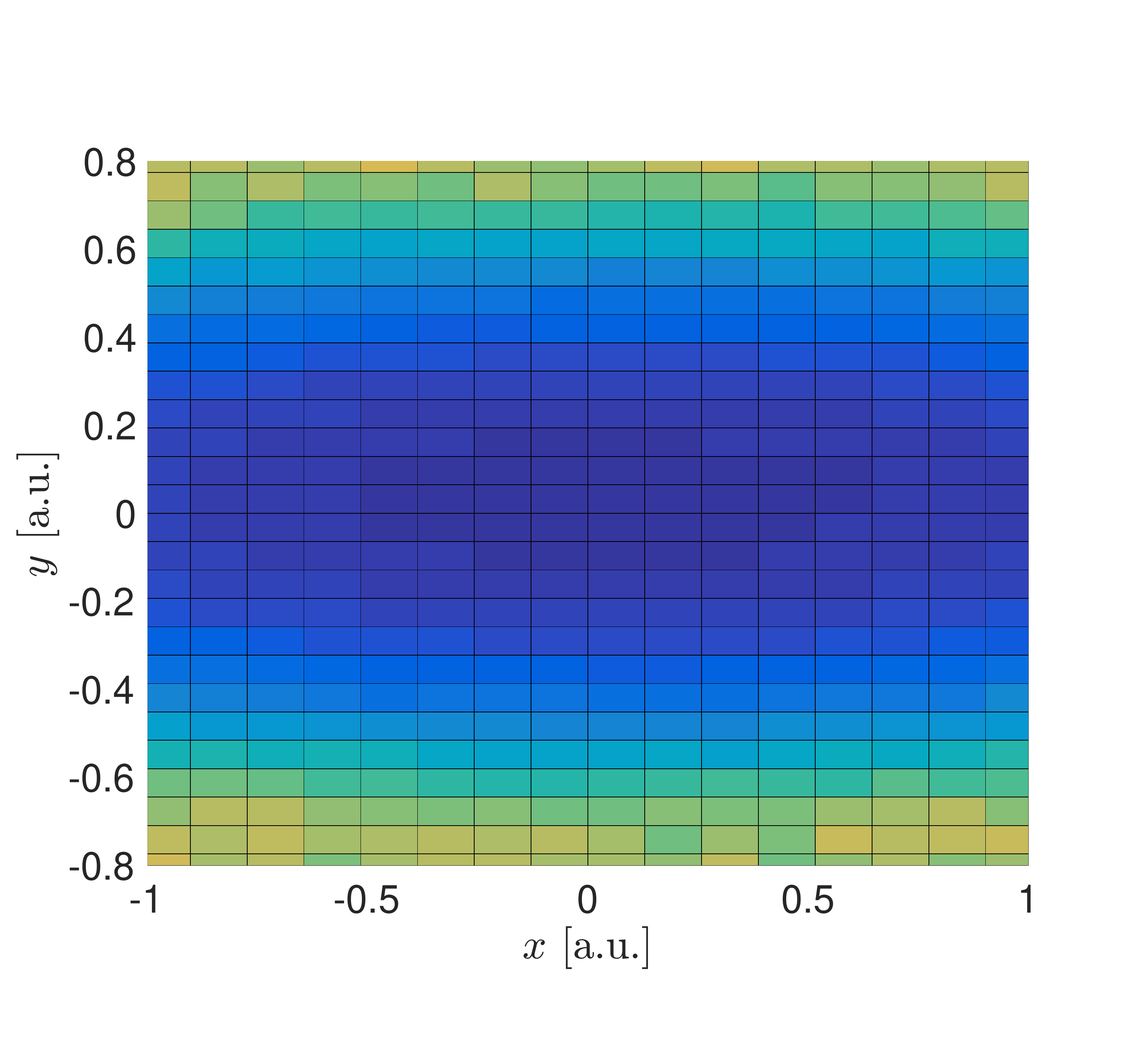} 
      \includegraphics[width=.32\columnwidth]{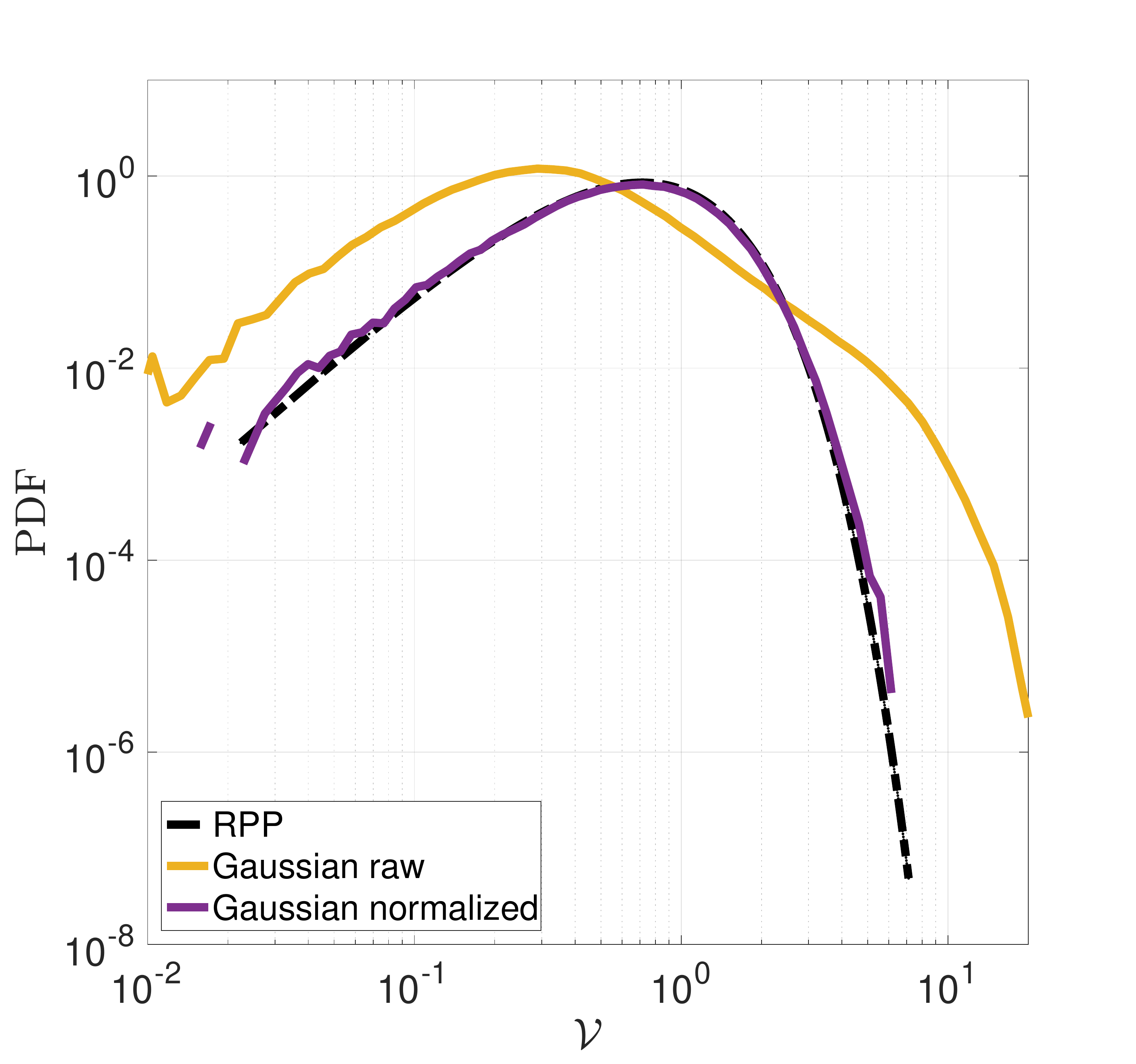}\\
	(a) \hspace{.33\textwidth} (b) \hspace{.33\textwidth}      (c)
        \caption{(a) Example of one realization of particles randomly distributed, with a large scale centered Gaussian modulation of the probability of presence (particles are more likely to be detected near the center than in the borders), mimicking the experimental illumination non-homogeneity due to the Gaussian profile of the laser sheet. (b) Coarse grained  field of the average local Vorono\"i area ${\cal A}_{mean}(x,y)$, estimated from 1000 realizations as in figure (a). (c) PDF of Vorono\"i areas, estimated from 1000 synthetic images (with a few hundreds of particles in each image) for:  (black dashed line) a random homogeneous RPP reference situation; (yellow solid line) a random but non-homogeneous distribution as illustrated in (a) and (purple solid line) the same random non-homogeneous distribution where Vorono\"i areas are locally corrected by the contraction field ${\cal A}_{mean}(0,0)/{\cal A}_{mean}(x,y)$ shown in (b).}
      \label{fig:RPPGaussian}
\end{figure*}

\subsection{Illumination inhomogeneity correction}
\begin{figure}[t]
\includegraphics[width=0.67\columnwidth]{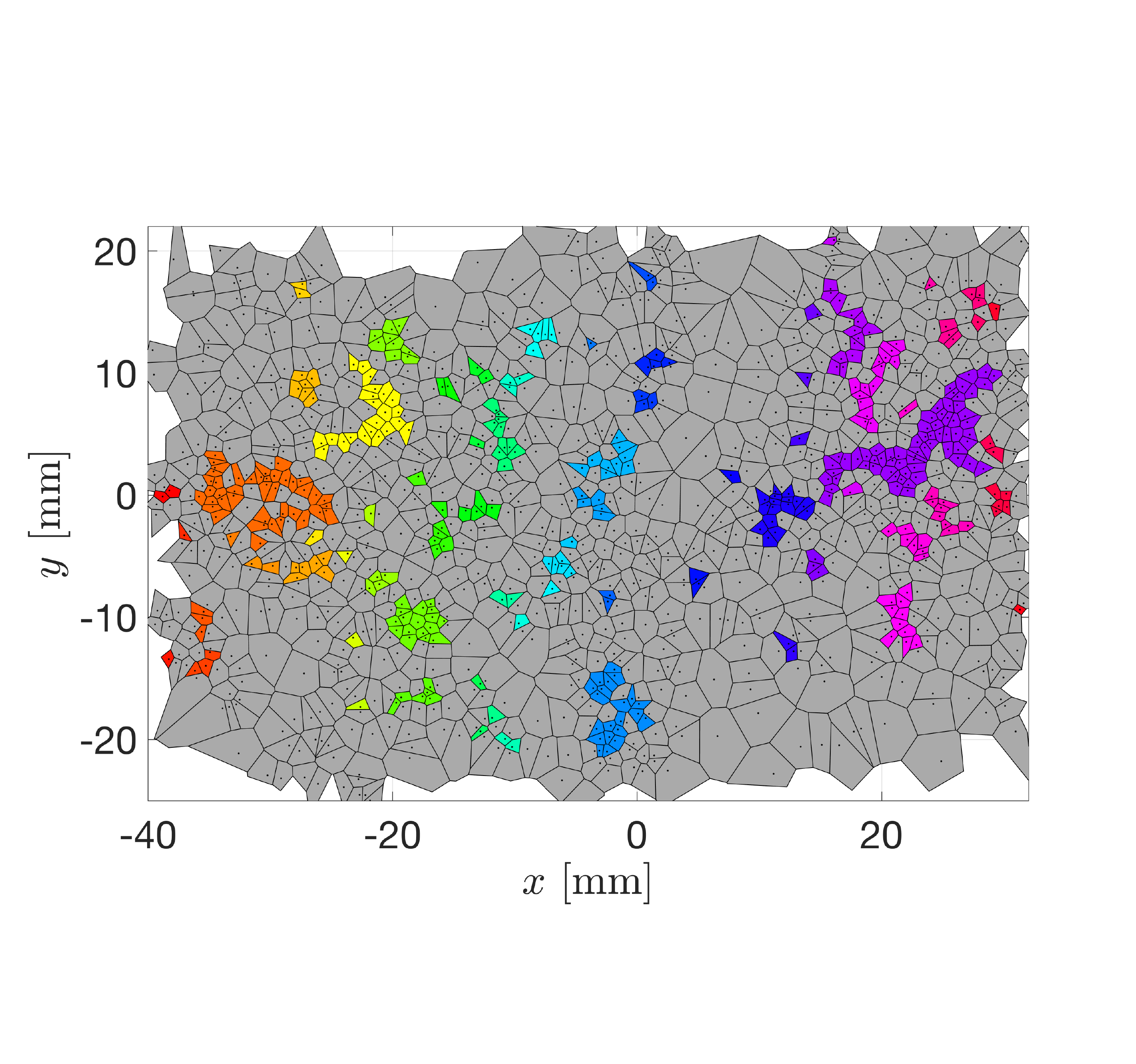}
\caption{Example of Vorono\"i diagram for one typical image of our experiment. Colored regions indicate clusters, defined following the procedure described in sec.~\ref{sec:clusters}}\label{fig:voroExp}
\end{figure}
Illumination inhomogeneity, shown above in figure~\ref{fig:raw_image}c, imposes additional image processing to calibrate the particle detection prior to diagnosing preferential concentration. As a consequence of the Gaussian intensity profile across the laser plane, particles are statistically more probable to be detected in the center of the visualization domain. Figure~\ref{fig:raw_image}a shows an example of a raw recorded image and figure~\ref{fig:raw_image}b indicates the corresponding particle detection. The map of probability of particle detection (figure~\ref{fig:raw_image}c), clearly shows that particles are more likely to be detected in the center of the image. Analyzing the clustering properties of particles in such conditions, without calibration, may lead to the errors in the diagnosis of the existence of clustering, simply because due to illumination issues. To prevent such a bias, previous studies have  cropped images~\cite{bib:obligado2015}, limiting the analysis to the central region, where illumination is relatively homogeneous. Doing so, however, requires many more images for statistical convergence of the analysis, and also biases the cluster/void analysis, as large structures cannot be detected. We use an alternative approach, allowing the use of the full image with an appropriate correction to undo the bias in the estimation of the area of Vorono\"i areas where illumination is non-homogeneous. A corrective local \emph{contraction} factor is applied to the raw Vorono\"i cells in regions with lower illumination to correct for them being statistically larger. We illustrate the method using a synthetically-generated random distribution of particles with a smooth gaussian modulation. Figure~\ref{fig:RPPGaussian}a represents one realization of the synthetically-generated particle field. Particles are randomly distributed following an RPP, but with a large-scale gaussian modulation mimicking the experimental bias in the center of the images. Figure~\ref{fig:RPPGaussian}c shows that, although no clustering mechanism is present, the PDF of normalized Vorono\"i areas ${\cal V}={\cal A \slash \langle \cal A \rangle}$ deviates significantly from the RPP case, simply because of the large scale modulation of the probability of particle location. The standard deviation of ${\cal V}$ is $\sigma_{\cal V}\simeq 1.5 > \sigma_{\cal V}^{RPP} =0.53$. To correct this bias, the coarse-grained field of the local average Vorono\"i area, ${\langle {\cal A} (x,y) \rangle \slash \langle {\cal A} (0,0) \rangle}$ (Figure~\ref{fig:RPPGaussian}b) is estimated from an ensemble of 1000 realizations. The color of each rectangular zone in Figure~\ref{fig:RPPGaussian}b represents the average value of the Vorono\"i area of particles detected within that zone. For smooth and large-scale inhomogeneities, the number of zones used for the coarse-grained field is not a critical parameter. This coarse-grained field is then used as a contraction factor, normalized to be maximum and equal to one where the particle probability is maximum, so that the Vorono\"i area ${\cal A}$ of a particle ${\cal P}$, detected at a position $(x,y)$ is corrected to become ${\cal A}^*={\cal A}\cdot \langle {\cal A} (0,0) \rangle \slash \langle {\cal A} (x,y) \rangle$ The PDF of the corrected Vorono\"i areas ${\cal V}*={\cal A}^*/\langle {{\cal A}^*} \rangle$, is shown in Figure~\ref{fig:RPPGaussian} and found to exactly match the reference RPP PDF, proving that the calibration with this correction method effectively removes the bias. The same procedure is used to unbias the  Vorono\"i area statistics from the experimental images.    

\section{Results}\label{sec:results}
\subsection{Deviation from Randomness of the Particle Concentration Field: Standard deviation ${\sigma}_{\nu}$ of the Vorono\"i area distributions}
\begin{figure*}[t]
  	\centering
      \includegraphics[height=6.3cm]{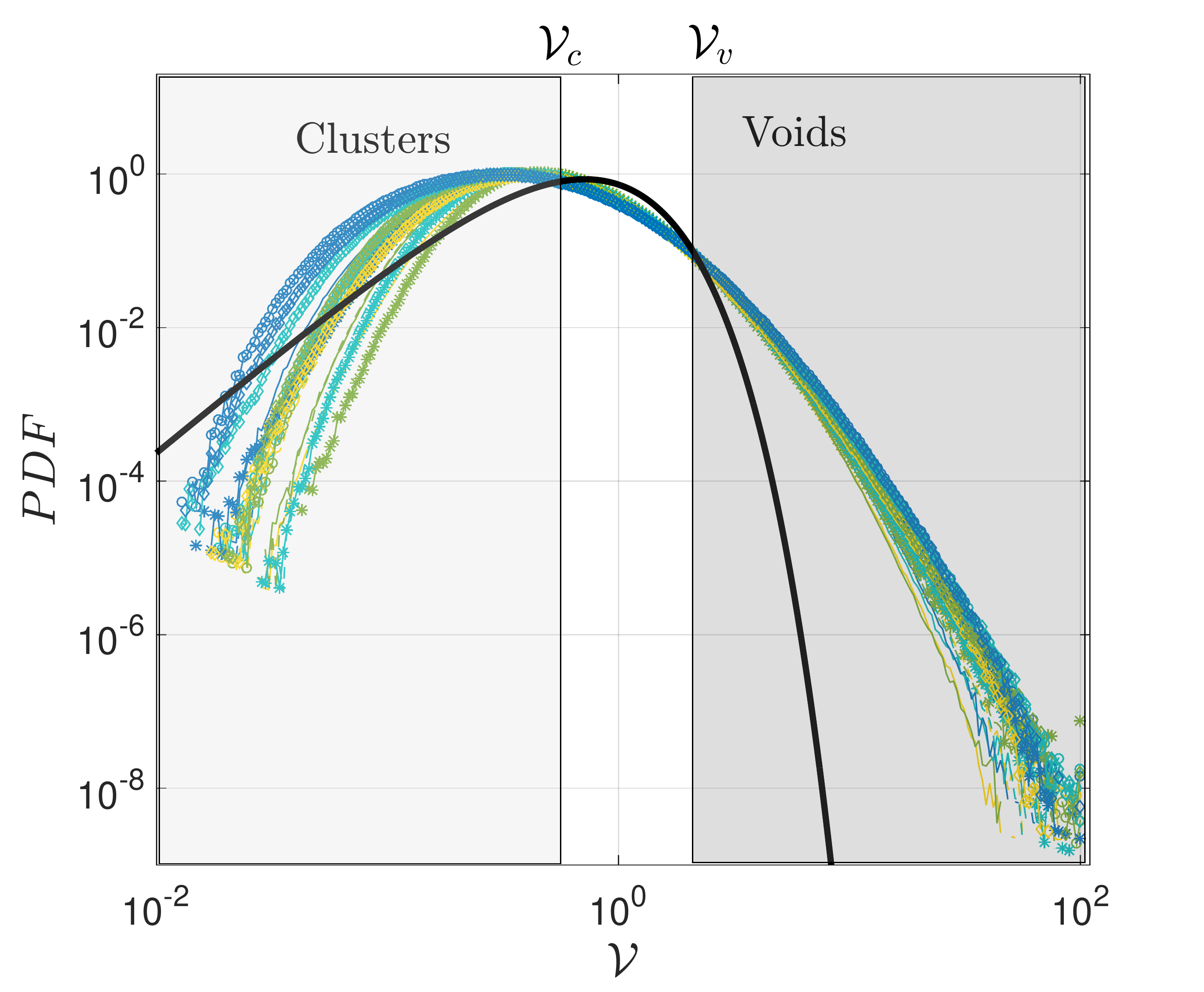}
      \includegraphics[height=6.3cm]{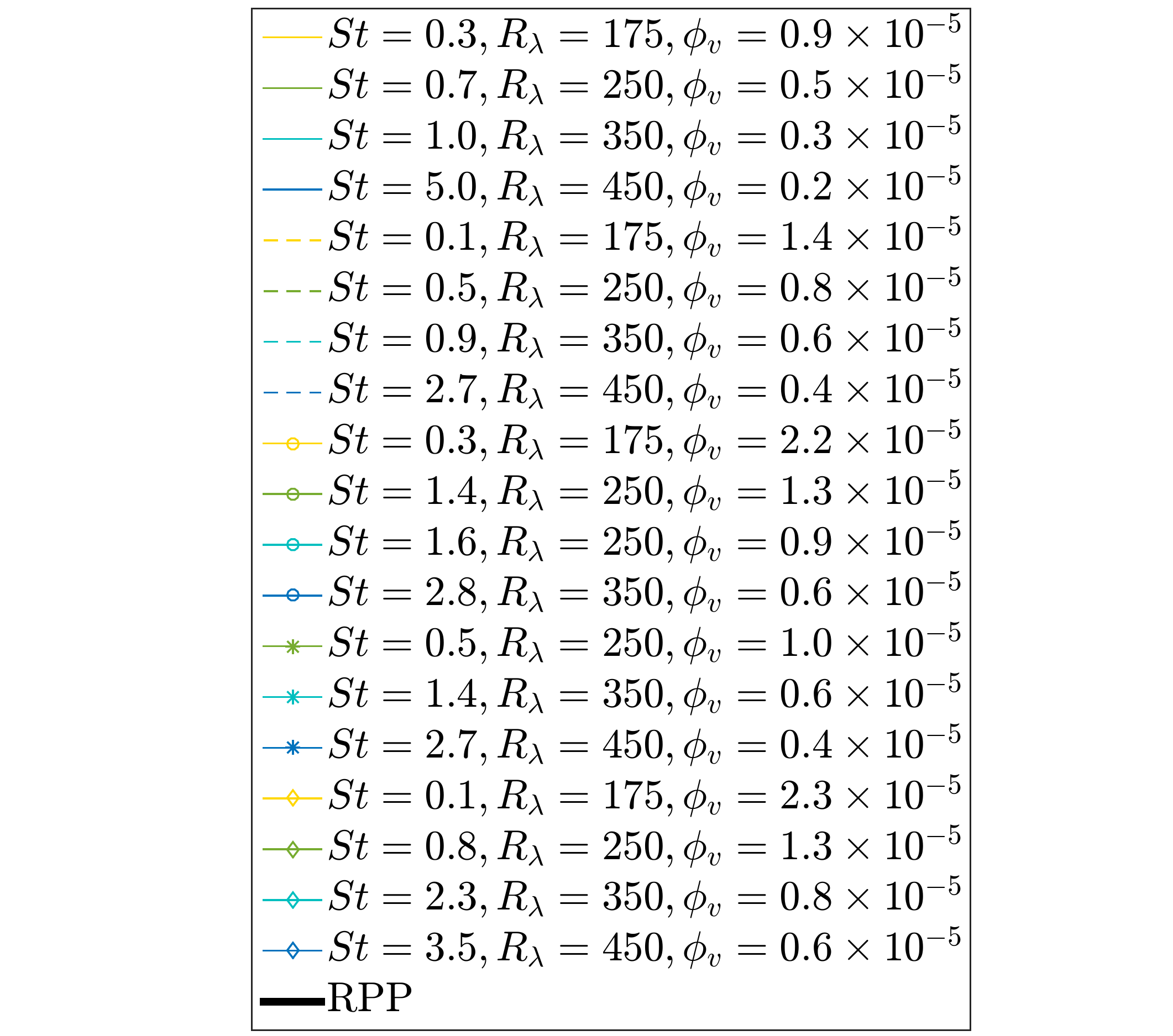} 
      \\
        (a) \hspace{.5\columnwidth} (b) \hspace{.5\columnwidth}    
      \includegraphics[height=6.3cm]{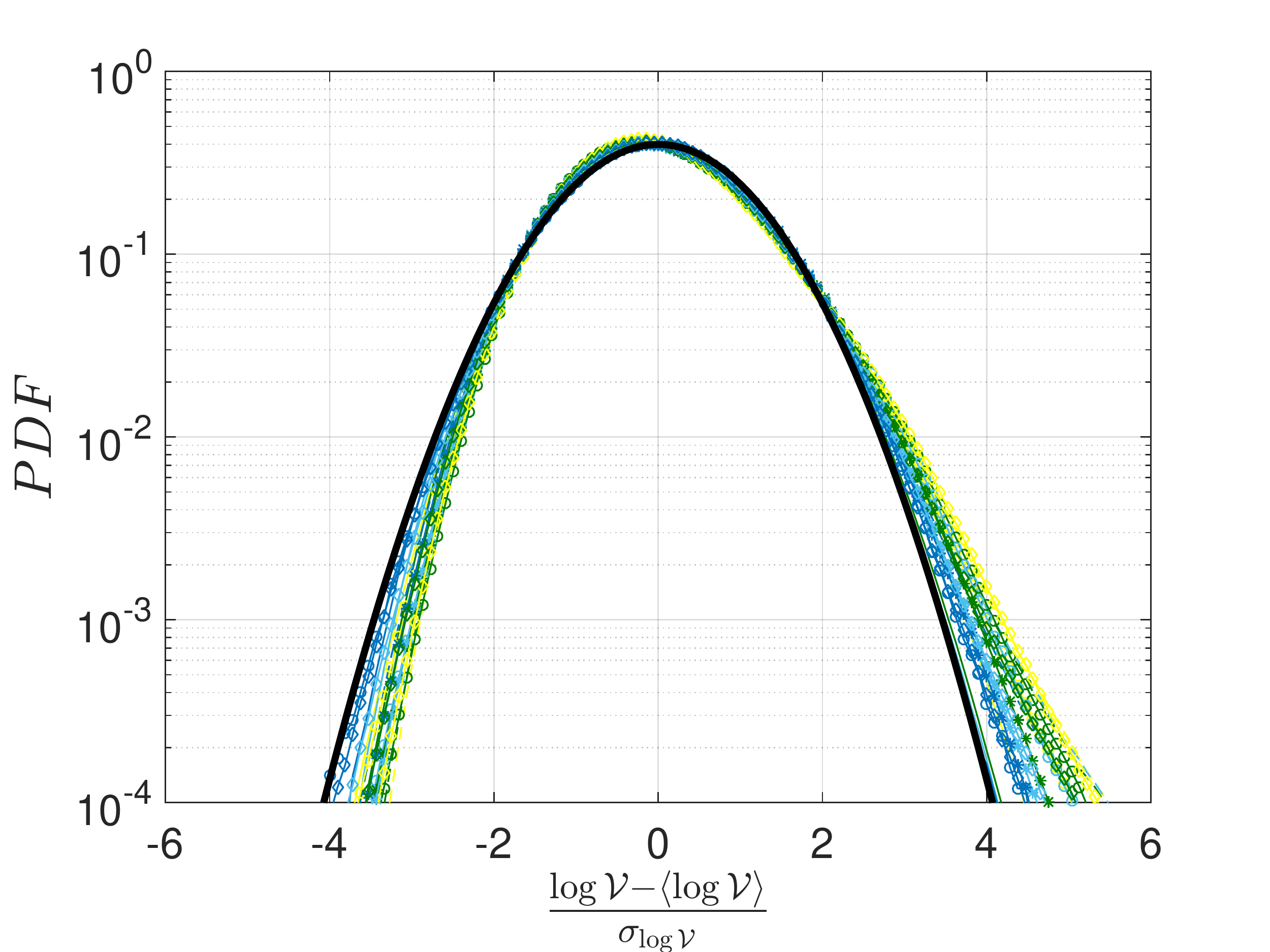}\\
      (c) \hspace{.5\columnwidth}
      \caption{(a) PDF of the corrected normalized Vorono\"i areas $\cal V$ for all experiments. The solid black line shows the RPP distribution. (b) Centered, normalized PDF of $\log({\cal V})$. The solid black line shows a gaussian distribution with variance 1.}
      \label{fig:voronoiiPDF}
\end{figure*}

Figure~\ref{fig:voroExp} represents a typical Vorono\"i diagram from an experimental image. Colored structures represent detected clusters, further discussed below. Thousands of such tesselations are obtained for each experimental condition. PDFs of normalized corrected Vorono\"i areas are shown in Figure~\ref{fig:voronoiiPDF}a, where the departure from the RPP case can be clearly seen. Figure~\ref{fig:voronoiiPDF}b shows the PDF of $\log(\cal V)$ (centered by the mean and normalized by the standard deviation), emphasizing the quasi log-normal distribution of the statistics of Vorono\"i areas, as  previously reported~\cite{bib:monchaux2010_PoF,bib:obligado2014_JoT}. This quasi-lognormality justifies the idea that the statistics of $\cal V$ can be described by a single parameter (recall that $\langle \cal V \rangle = 1 $ by construction), generally the standard deviation of ${\cal V}$, $\sigma_{\cal V}$, to quantify the departure from the RPP distribution. 

\begin{figure}[h!] 
  	\centering
      \includegraphics[width=.49\columnwidth]{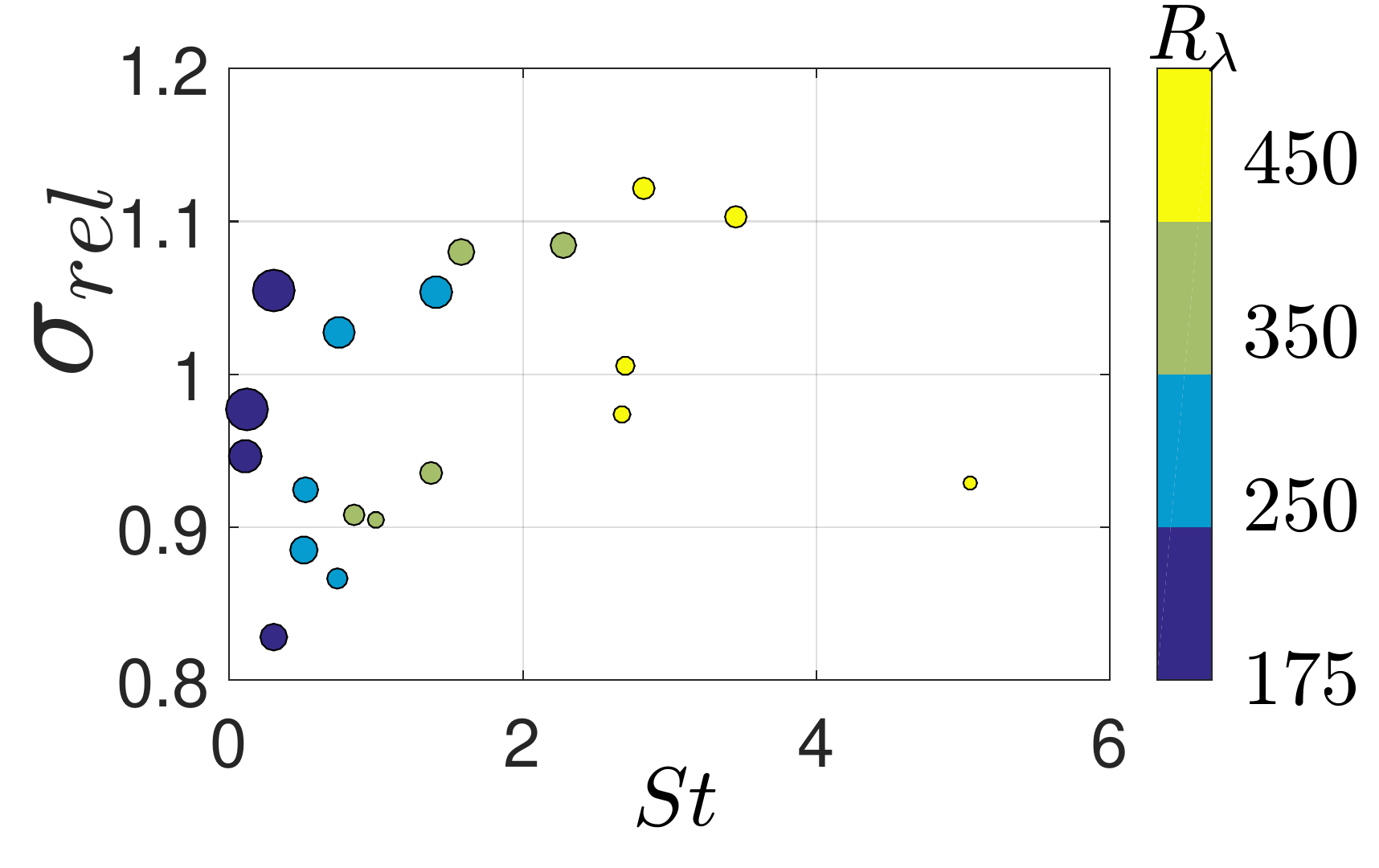}
      \includegraphics[width=.49\columnwidth]{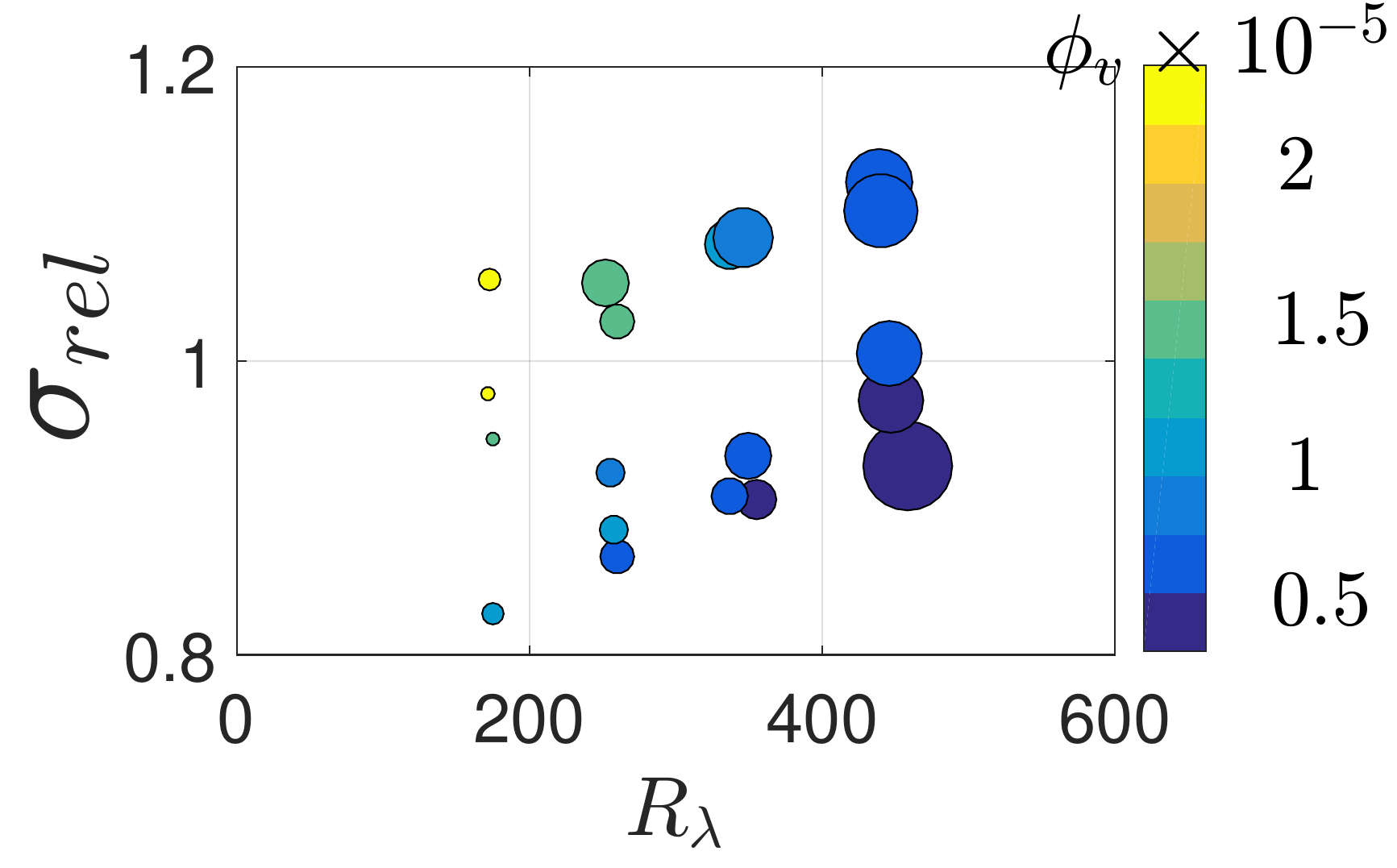}\\
     (a) \hspace{0.5\columnwidth} (b) \\
      \includegraphics[width=.49\columnwidth]{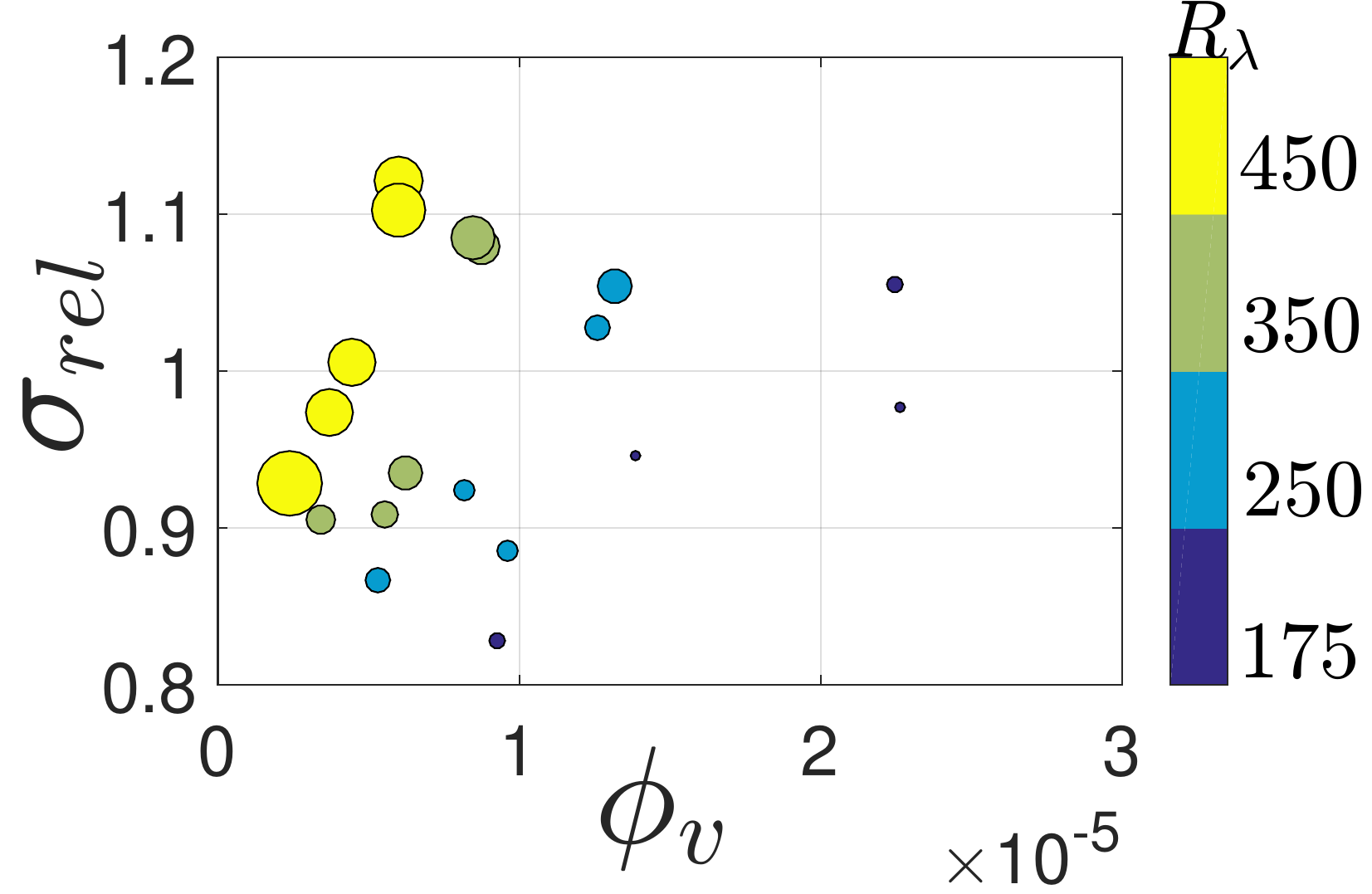} \\
      (c)
      \caption{(a) Standard deviation $\sigma$ vs $St$. Symbol colors represent the Reynolds number,  while the size of the symbols encodes the volume fraction (larger symbols correspond to experiments at larger volume fraction). (b) Standard deviation $\sigma$ vs $Re_{\lambda}$. Symbol colors indicate the volume fraction, while the size of the symbols encodes the Stokes number (larger symbols correspond to experiments with larger Stokes numbers). (c) Standard deviation $\sigma$ vs $\phi_v$. Symbol colors reflect the Reynolds number, while the size of symbols encodes the Stokes number (larger symbols correspond to experiments with particles at larger Stokes number).}
      \label{fig:sigma_vs_StRePhi}
 \end{figure}
 
\begin{figure} 
  	\centering
      \includegraphics[width=.49\columnwidth]{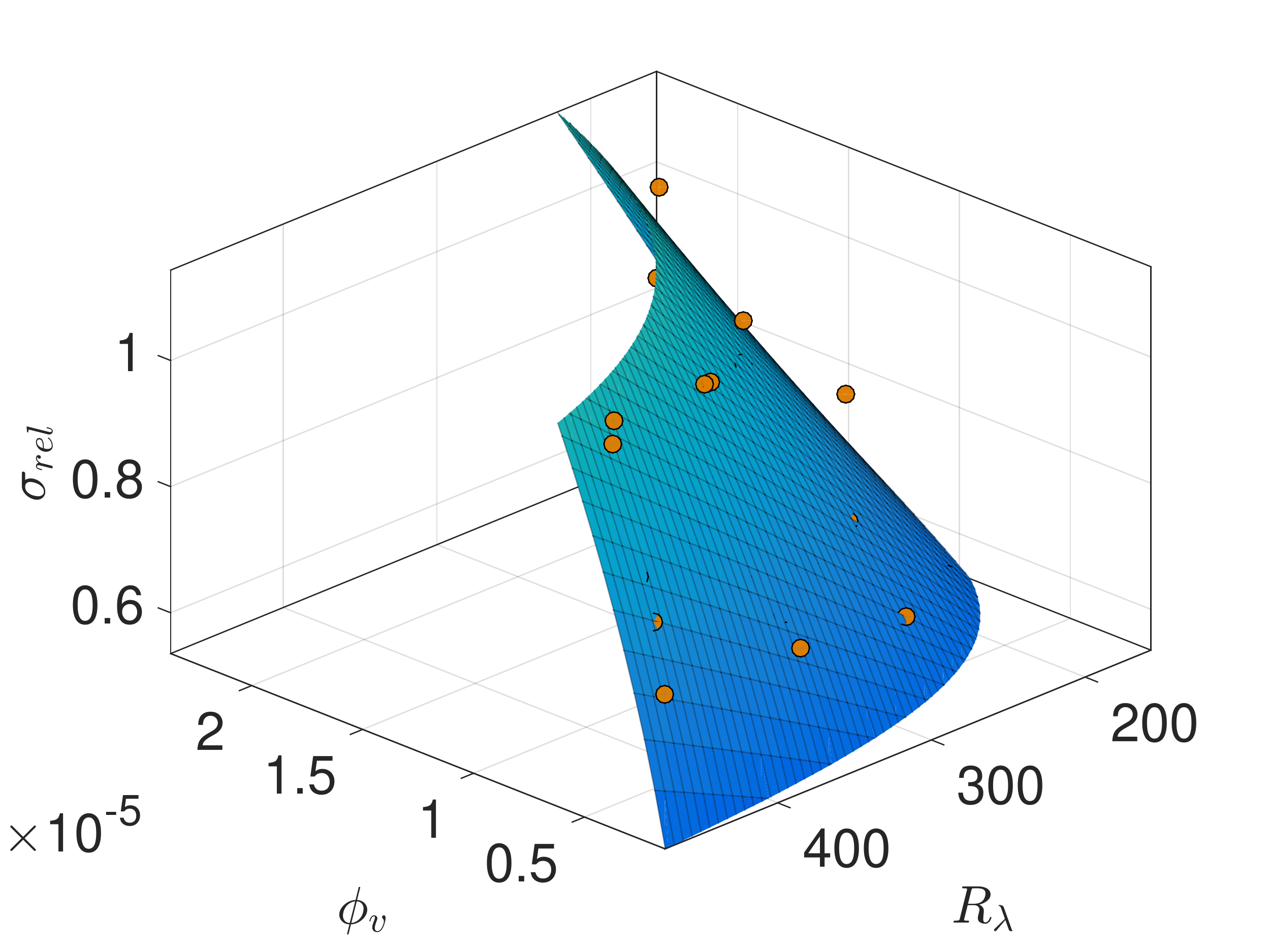}
      \includegraphics[width=.49\columnwidth]{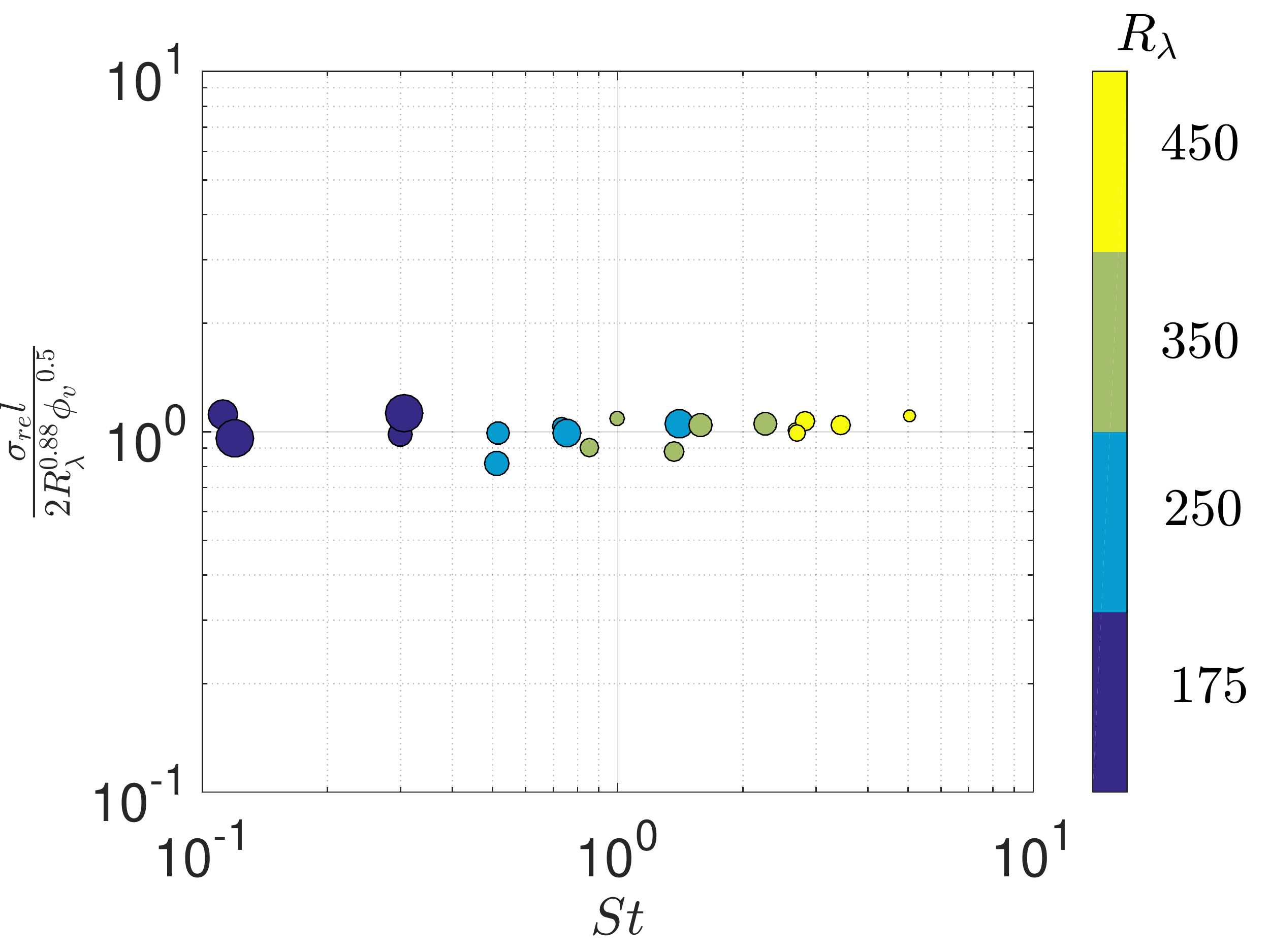}\\
      (a) \hspace{.5\textwidth} (b)
      \caption{(a) $\sigma_{rel}$ as a function of $\phi_v$ and $Re_\lambda$ with a power law fit $\sigma_{rel}\propto Re_\lambda^{\beta}\phi_v^{\gamma}$.  (b) $\sigma_{rel}$ compensated by $Re_\lambda^{\beta}\phi_v^{\gamma}$. Best fit is obtained for $\beta \simeq 0.88$ and $\gamma \simeq 0.5$.}
      \label{fig:sigma_scaling}
      \end{figure}    
      
Figure~\ref{fig:sigma_vs_StRePhi} represents the difference between the experimental $\sigma_{\cal V}$ and the RPP value ($\sigma_{rel}=\left(\sigma_{\cal V}-\sigma^{RPP}_{\cal V}\right)/\sigma^{RPP}_{\cal V}$) as a function of Stokes number, $Re_\lambda$ and $\phi_v$. It is found that, for all experiments, $\sigma_{rel} > 0$, consistent with the existence of clustering. No clear trend with $St$ can be identified.  The most striking observation from these figures is the strong dependency of $\sigma_{\cal V}$ on the volume fraction. This is highlighted in figure~\ref{fig:sigma_vs_StRePhi}c where, for every Reynolds number, $\sigma_{rel}$ is observed to increase quasi-linearly with $\phi_v$. Trends with Reynolds number are more difficult to extract from this simple projection, although figure~\ref{fig:sigma_vs_StRePhi}a, where the Reynolds number dependency is encoded in the color of the symbols, suggests  an increase of $\sigma_{\cal V}$ with $Re_\lambda$.


To further quantify the dependencies of $\sigma_{rel}$ with the three control parameters $(St,Re_\lambda,\phi_v)$, power law fits are computed from the entire experimentally-sampled spaced:

\begin{equation}
\sigma_{rel}=K St^\alpha Re_\lambda^\beta \phi_v^\gamma,
\end{equation}

Based on the observations from figure~\ref{fig:sigma_vs_StRePhi}, we first determine $\beta$ and $\gamma$ by a two-variables fit of the $\sigma_{rel}$ data as a function of $Re_\lambda$ and $\phi_v$, neglecting the dependency on $St$ in a first approximation. This is partially supported by the lack of a consistent evolution with Stokes number on figure~~\ref{fig:sigma_vs_StRePhi}a. The corresponding data and fit are shown in fig.~\ref{fig:sigma_scaling}a, where the best fit is obtained for $\beta\simeq 0.88\pm 0.2$ and $\gamma \simeq 0.5\pm 0.15$. The dependency of $\sigma_{rel}$ with $St$ is then explored by plotting the compensated quantity $\frac{\sigma_{rel}}{Re_\lambda^\beta\phi_v^\gamma}$,  as a function of $St$ in figure~\ref{fig:sigma_scaling}b. The data points present very little scatter, with no  trend observed (best power law fit results in a exponent $\alpha\simeq0.0\pm 0.05$).  
Overall, the dependency of the standard deviation of the Vorono\"i area distribution with the three controlling parameters $(St,Re_\lambda,\phi_v)$ results in the empirical scaling:
\begin{equation}
\sigma_{rel}=\frac{\sigma_{\cal V}-\sigma_{\cal V}^{RPP}}{\sigma_{\cal V}^{RPP}}\simeq 2 \; St^{0.0}Re_\lambda^{0.88}\phi_v^{0.5}.
\end{equation}

Interestingly, our results point towards a dominant dependency of the clustering on the turbulent Reynolds number, with a smaller dependency on volume fraction and no dependency on Stokes number. 

\subsection{Contribution of Clusters and Voids to the Standard Deviation of the Vorono\"i Area Distribution}

We define clusters and voids in fig.~\ref{fig:voronoiiPDF}a, from the thresholds ${\cal V}_c$ and ${\cal V}_v$~\cite{bib:monchaux2010_PoF,bib:monchaux2012_IJMF}, corresponding to the points where the experimental Vorono\"i area PDF is above (more probable than) the RPP. Clusters are defined as particle ensembles with adjacent Vorono\"i cells whose area ${\cal V}<{\cal V}_c$ while voids correspond to cells whose area ${\cal V}>{\cal V}_v$. In the experiments reported here, the two cutoffs are insensitive to flow conditions, and their values, ${\cal V}_c=0.6$ and ${\cal V}_v=2.1$, are equal to those in previous studies at lower turbulent Reynolds numbers~\cite{bib:obligado2014_JoT,bib:obligado2015_EPL}. The invariance of these intersections remains to be understood. 

The standard deviation $\sigma_{\cal V}$ of Vorono\"i areas represents the second moment of the PDF of $\cal{V}$. One can therefore argue that large areas (\emph{i.e.} voids) contribute more to $\sigma_{\cal V}$ than small areas (\emph{i.e.} clusters). We can indeed write $\sigma_{\cal V}^2$ as
\begin{equation}
\sigma_{\cal V}^2=\int_0^{{\cal V}_c} \left({\cal V}-\bar{{\cal V}}\right)^2PDF({\cal V})\textrm{d}{\cal V} + \int_{{\cal V}_c}^{{\cal V}_v} \left({\cal V}-\bar{{\cal V}}\right)^2PDF({\cal V})\textrm{d}{\cal V}+\int_{{\cal V}_v}^\infty \left({\cal V}-\bar{{\cal V}}\right)^2PDF({\cal V})\textrm{d}{\cal V},
\end{equation}

where the three terms give the contribution of clusters, intermediate areas and voids (denoted as $\sigma_c$, $\sigma_i$ and $\sigma_v$), respectively, to the total standard deviation of Vorono\"i areas. 
For the RPP, the three contributions are comparable: $\sigma_{c}^{RPP} = 0.29$, $\sigma_{v}^{RPP} = 0.30$ and $\sigma_i^{RPP} = 0.32$. Obviously ${\sigma_c^{RPP}}^2+{\sigma_i^{RPP}}^2+{\sigma_v^{RPP}}^2 = 0.53^2$ as expected. The questions are: how do these contributions change for inertial particles and how do they evolve with the controlling parameters? The experimental data shows  that $\sigma_v^2$ represents on average $\approx75\%$ ($69\%-78\%$, depending on the experimental conditions) of the total variance $\sigma_{\cal V}^2$, $\sigma_c$ is only $\approx 17\%$  ($14\%-18\%$) and $\sigma_i^2 \approx 8\%$ ($6\%-12\%$). This partition clearly shows a stronger contribution of voids   to the total variance compared to clusters, in contrast to the random case, and as expected from the extended tails of the inertial particle Vorono\"i PDF. The standard deviation of Vorono\"i areas, as  commonly discussed in particle preferential accumulation, is therefore essentially a measure of the distribution of voids. From this point of view, we have analyzed how each of the three contributions, clusters, voids and intermediate areas, evolve with flow parameters. For each contribution, the relative deviation is compared to the RPP case($\sigma_{rel,*} = \frac{\sigma_{*}-\sigma_*^{RPP}}{\sigma_*^{RPP}}$, with $*=c, i$ or $v$) and their dependencies on Reynolds, Stokes numbers and volume fraction are:

\begin{equation}
	\sigma_{rel,c}=0.33 \times {Re_\lambda}^{1.05 \pm 0.44} \times {\phi_v}^{0.5 \pm 0.26} {St}^{0.01 \pm 0.07},  
\end{equation}
\begin{equation}
	\sigma_{rel,v}=4.39 \times {Re_\lambda}^{0.79 \pm 0.18} \times {\phi_v}^{0.46 \pm 0.11} {St}^{0.01 \pm 0.03}, 
\end{equation}
\begin{equation}
	\sigma_{rel,i}=0.35 \times {Re_\lambda}^{0.66 \pm 0.17} \times {\phi_v}^{0.41 \pm 0.09} {St}^{0.00 \pm 0.03}, 
\end{equation}

These power law fits show that although the strongest contribution comes indeed from the voids, the dependencies with experimental parameters are comparable for all zones, with a leading role for the Reynolds number, a lesser influence of the volume fraction and practically no dependency on Stokes number, within the range of explored parameters.




\subsection{Geometry of Clusters and Voids in the  Particle Concentration Field}\label{sec:clusters}


    

\begin{figure}[t]
  	\centering
      \includegraphics[width=.49\columnwidth]{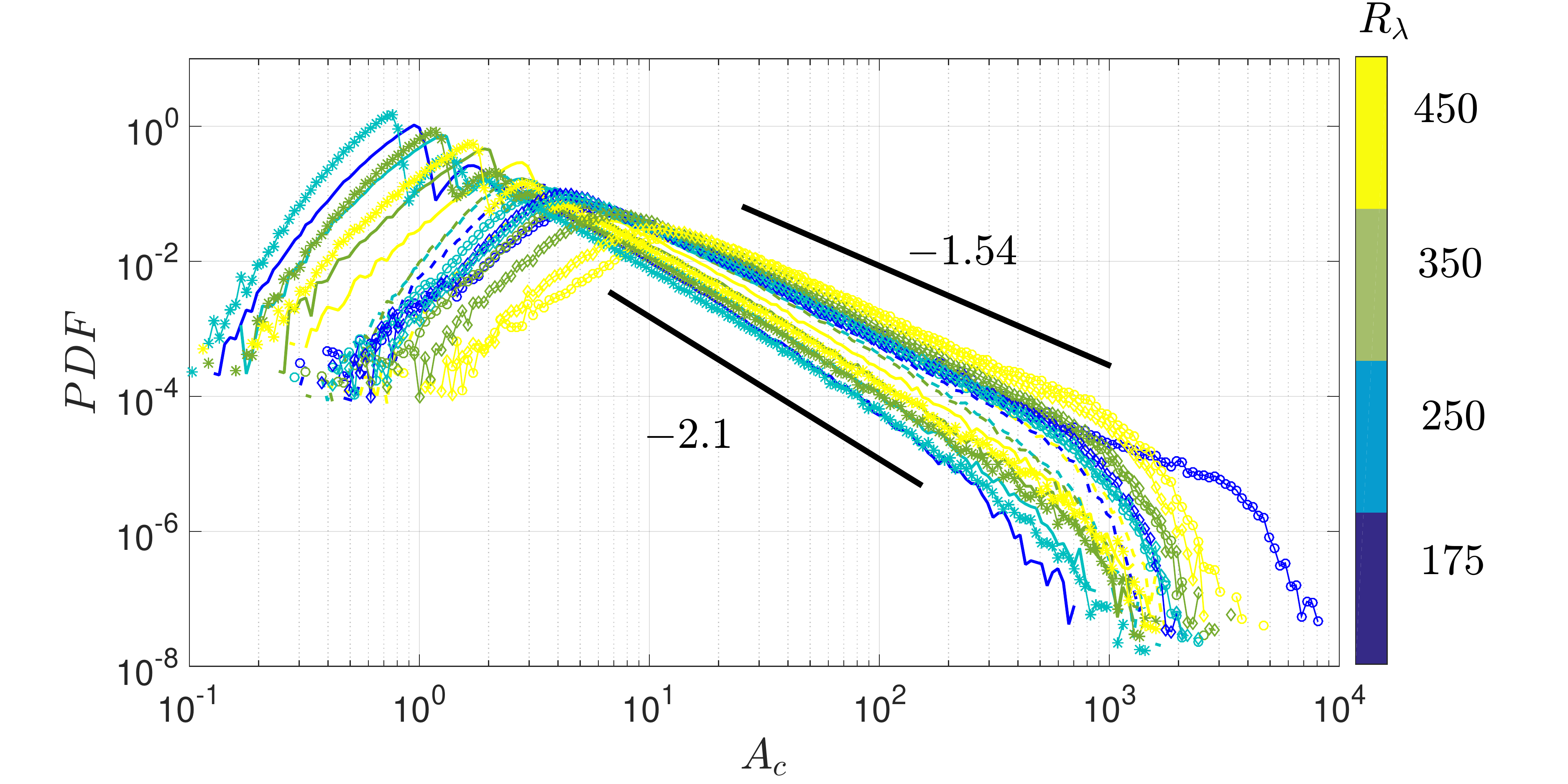}
      \includegraphics[width=.49\columnwidth]{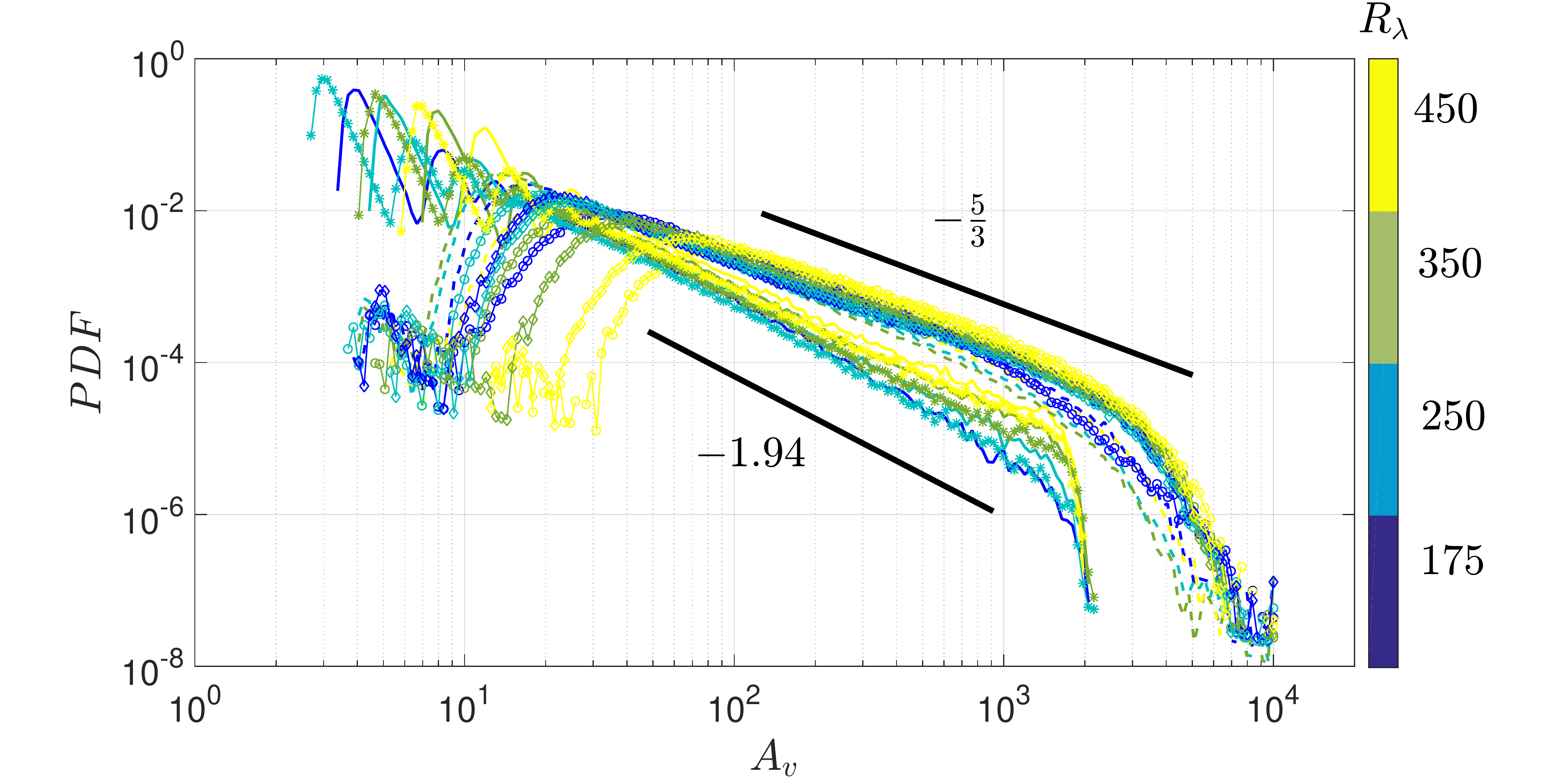}\\
       (a) \hspace{.5\columnwidth} (b)
       \includegraphics[width=.49\columnwidth]{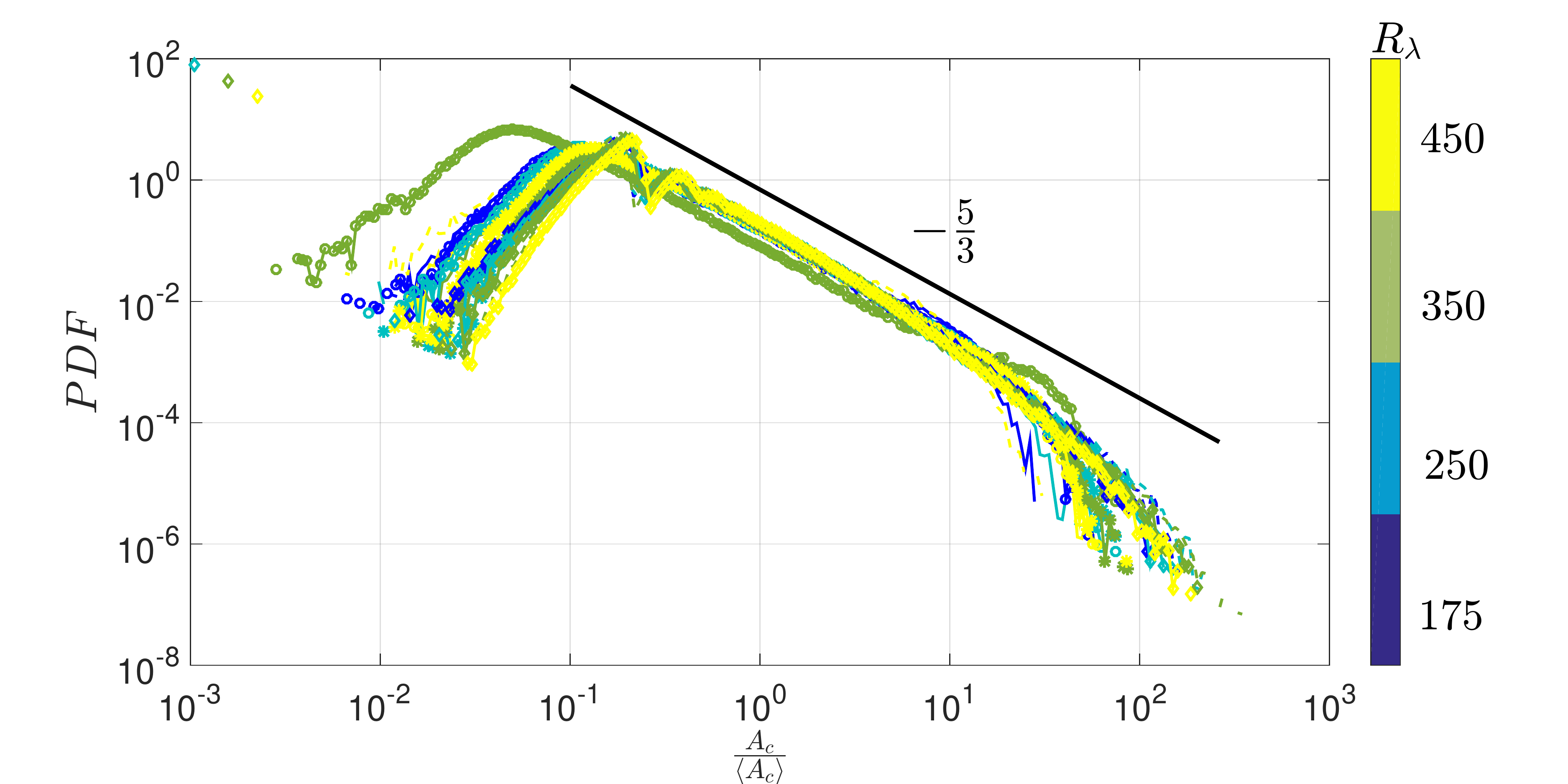}
      \includegraphics[width=.49\columnwidth]{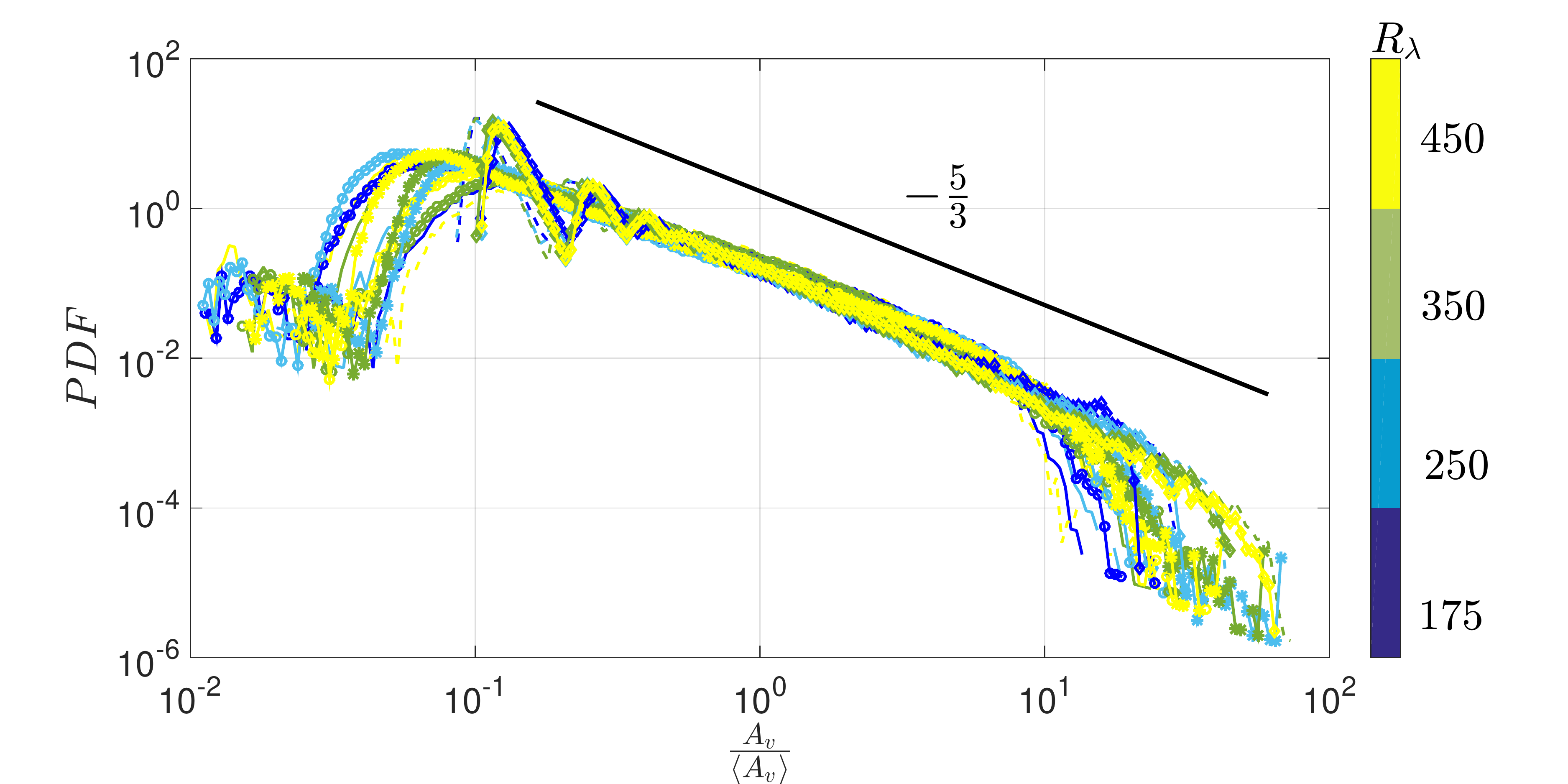}\\
       (c) \hspace{.5\columnwidth} (d)
      \caption{PDFs of clusters areas ${A_c}$ (a) and voids areas ${A_v}$ (b). Figs. (c) and (d) show the same PDFs for the areas normalized by the mean.}  
      \label{fig:clusters_area_voids} 
\end{figure}
 
Figure~\ref{fig:clusters_area_voids} presents the PDF of cluster and void areas, before (top) and after (bottom) normalization. The cluster PDFs exhibit a distinct peak, indicating the existence of a typical characteristic cluster dimension, in agreement with other previous experimental findings~\cite{bib:aliseda2002_JFM,Bateson12,bib:obligado2014_JoT,bib:obligado2015_EPL}. Figure~\ref{fig:clusters_area_voids}c shows that the PDFs of the normalized cluster areas $A_c/<A_c>$ follow an algebraic decay with an exponent $n_c\approx -5/3$, for areas larger than the most probable value. Similar trends are observed for the void areas PDFs, although the range of sizes of the voids is naturally larger than that of the clusters (by a factor about $10$). The exponent $n_v$ for the decay of the PDF of normalized void area follows a similar trend to the clusters ($n_v\approx -5/3$). These qualitative features are found to be robust for all experimental conditions. Algebraic decay of the cluster and void areas have been previously reported in several previous experimental and numerical studies~\cite{bib:boffetta2004_PoF,bib:goto2006_PoF,bib:monchaux2010_PoF,bib:obligado2014_JoT,bib:obligado2015_EPL} and is in agreement with a simple model proposed in~\cite{bib:goto2006_PoF} which predicts an algebraic decay for the PDF of void areas with a $-5/3$ exponent. In this model, the distribution of voids mimics the self-similar distribution of eddies across the turbulent energy cascade, suggesting that clustering (and voiding) of inertial particles is not only driven by small scales but reflects the self-similarity of the carrier turbulence. Unlike in the original work proposing the model, where it applied across the entire spectrum (from $\eta$ to $L_{int}$), in these experiments, the $-5/3$ decay holds between a lower length scale comprised between $3$ and $10 \eta$, depending on flow conditions, and an upper length scale slightly below $L_{int}$. The largest length scales are not fully resolved in the experiments since the images are about $L_{int}$, so the tails in the right hand side of the distributions (Figure~\ref{fig:clusters_area_voids}) are not statistically significant.

\begin{figure}[t]
  	\centering
      \includegraphics[width=.49\columnwidth]{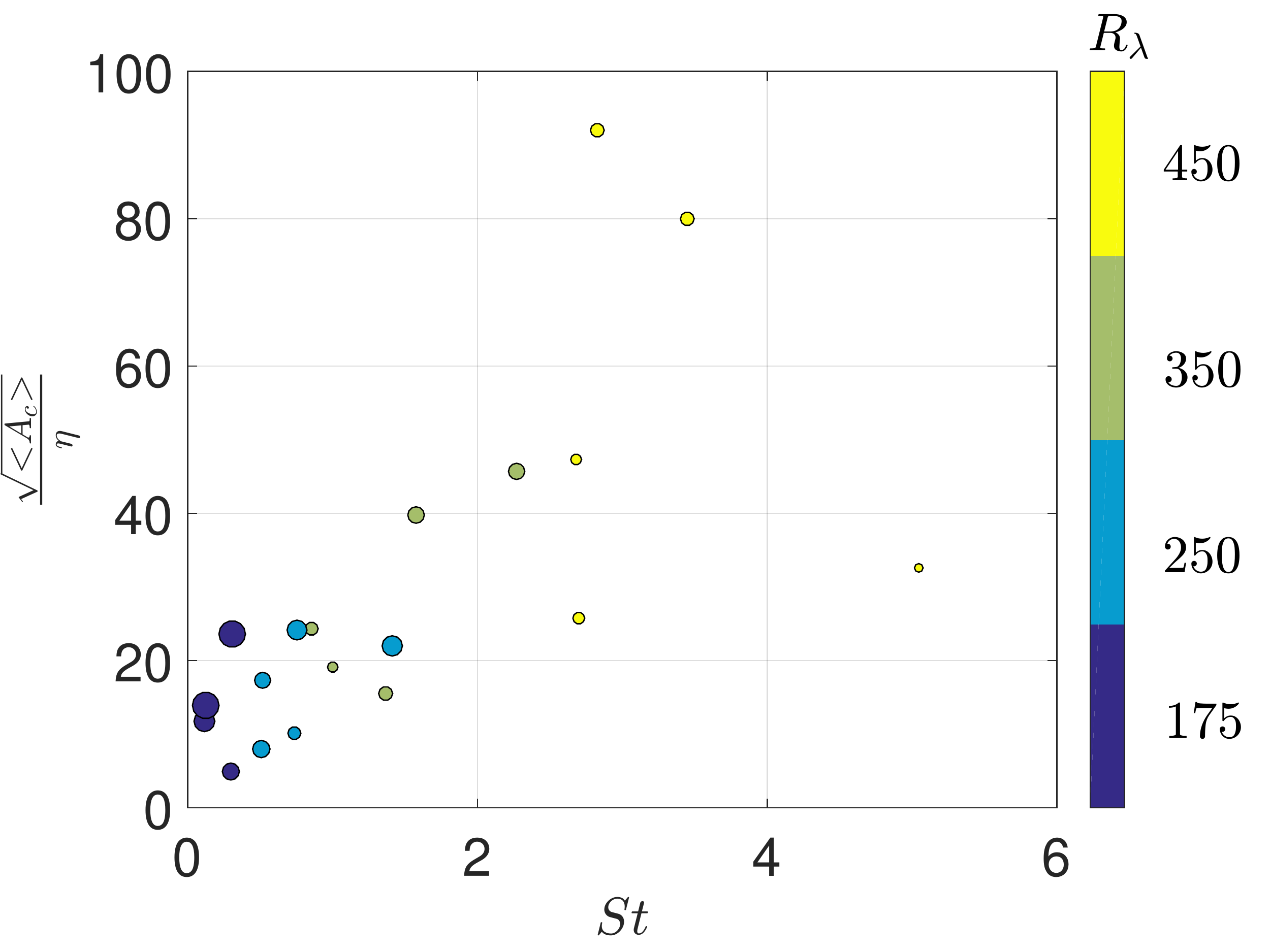}
      \includegraphics[width=.49\columnwidth]{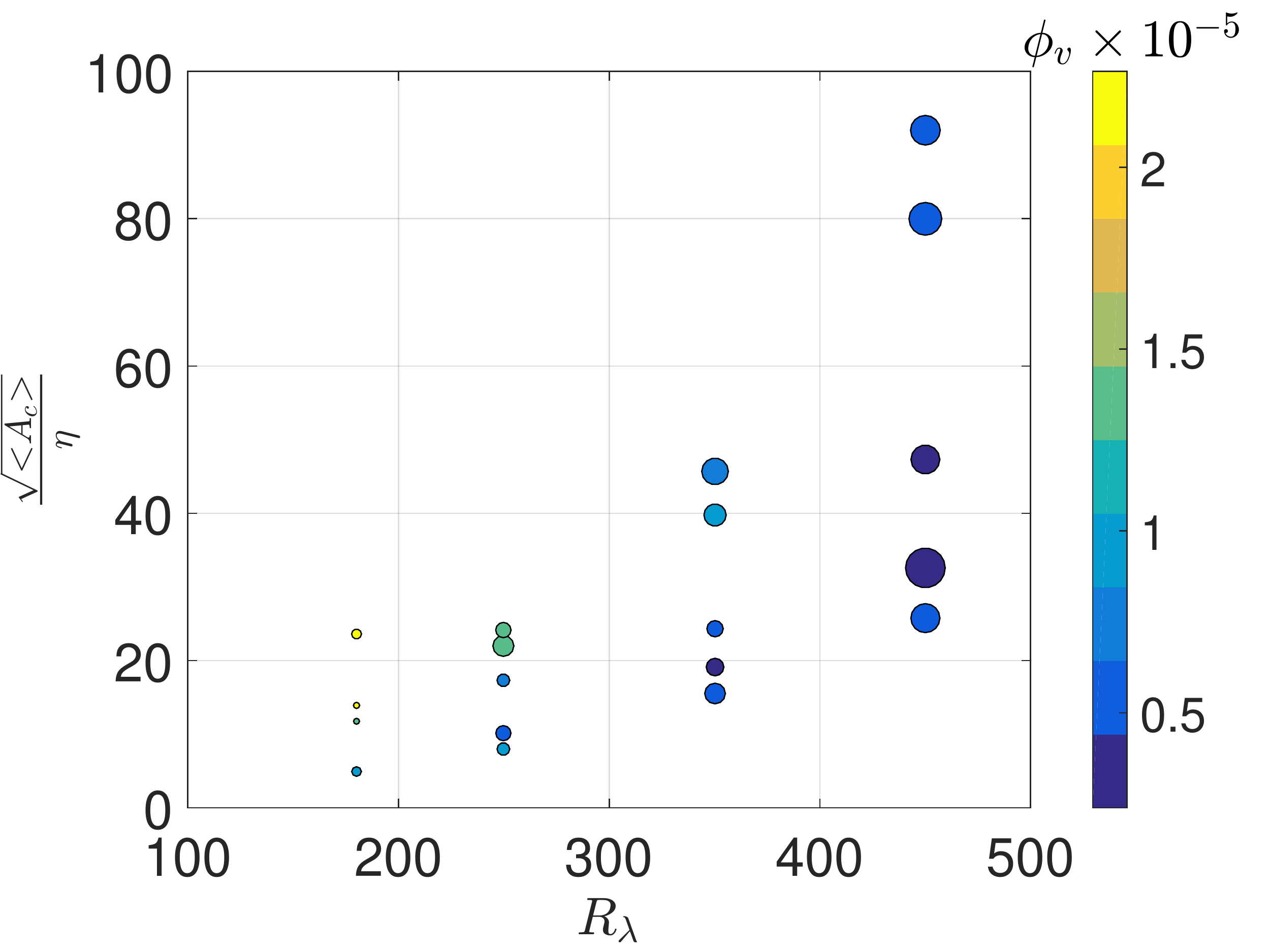}\\
      (a) \hspace{.5\columnwidth} (b)\\
      \includegraphics[width=.49\columnwidth]{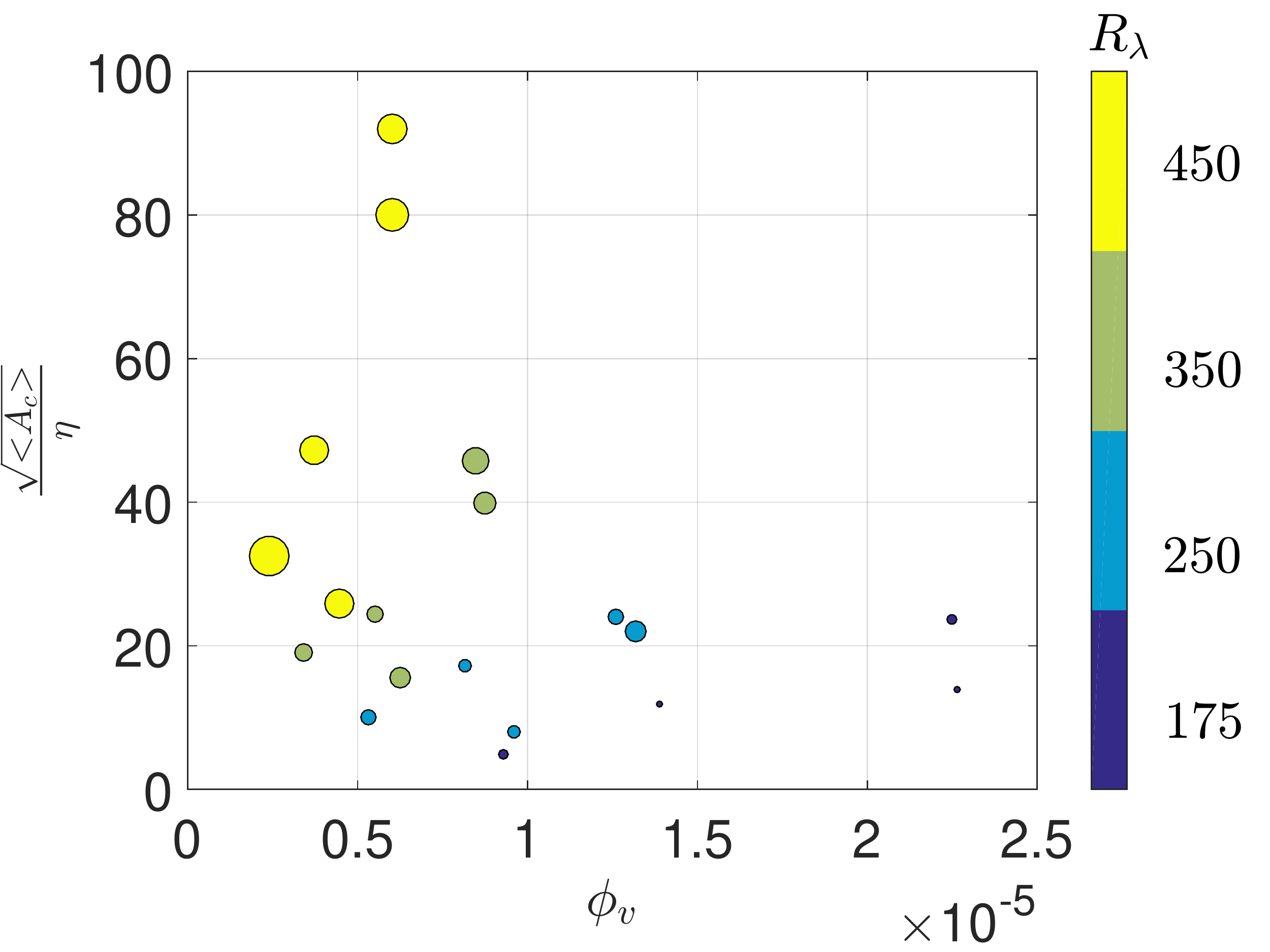}\\
      (c)
      \caption{Dependency of the average cluster size on Stokes number (a), Reynolds number (b) and volume fraction (c). In (a), the color of the symbols indicates the Reynolds number and the size of the symbols reflects the volume fraction. In (b), the color of the symbols indicates the volume fraction and the size of the symbols reflects the Stokes number. In (c), the color of the symbols indicates the Reynolds number and the size of the symbols reflects the Stokes number.}
      \label{fig:clusters_StRePhi}
\end{figure} 

\begin{figure}[t]
	\centering
      \includegraphics[width=.49\columnwidth]{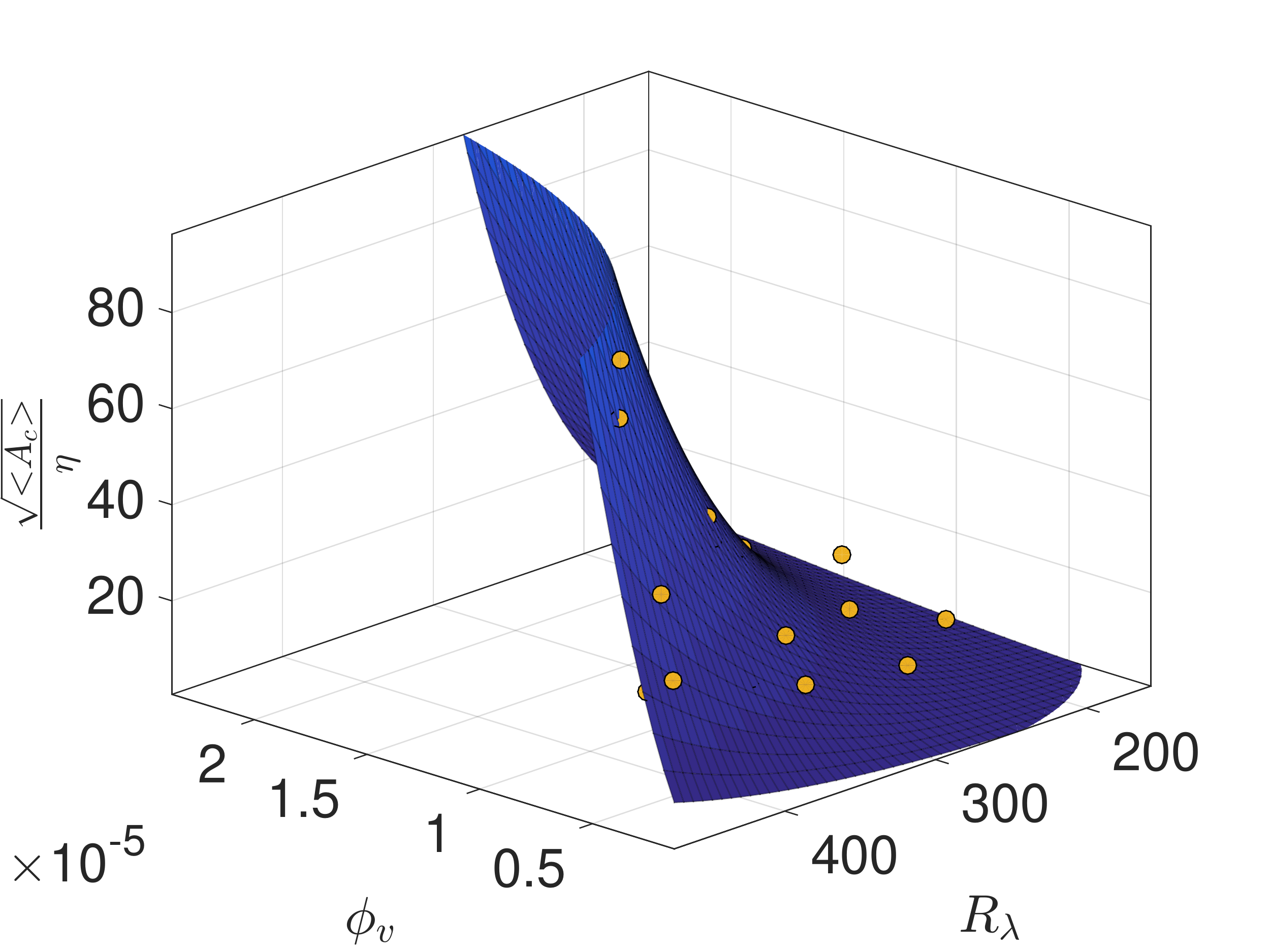}
      \includegraphics[width=.49\columnwidth]{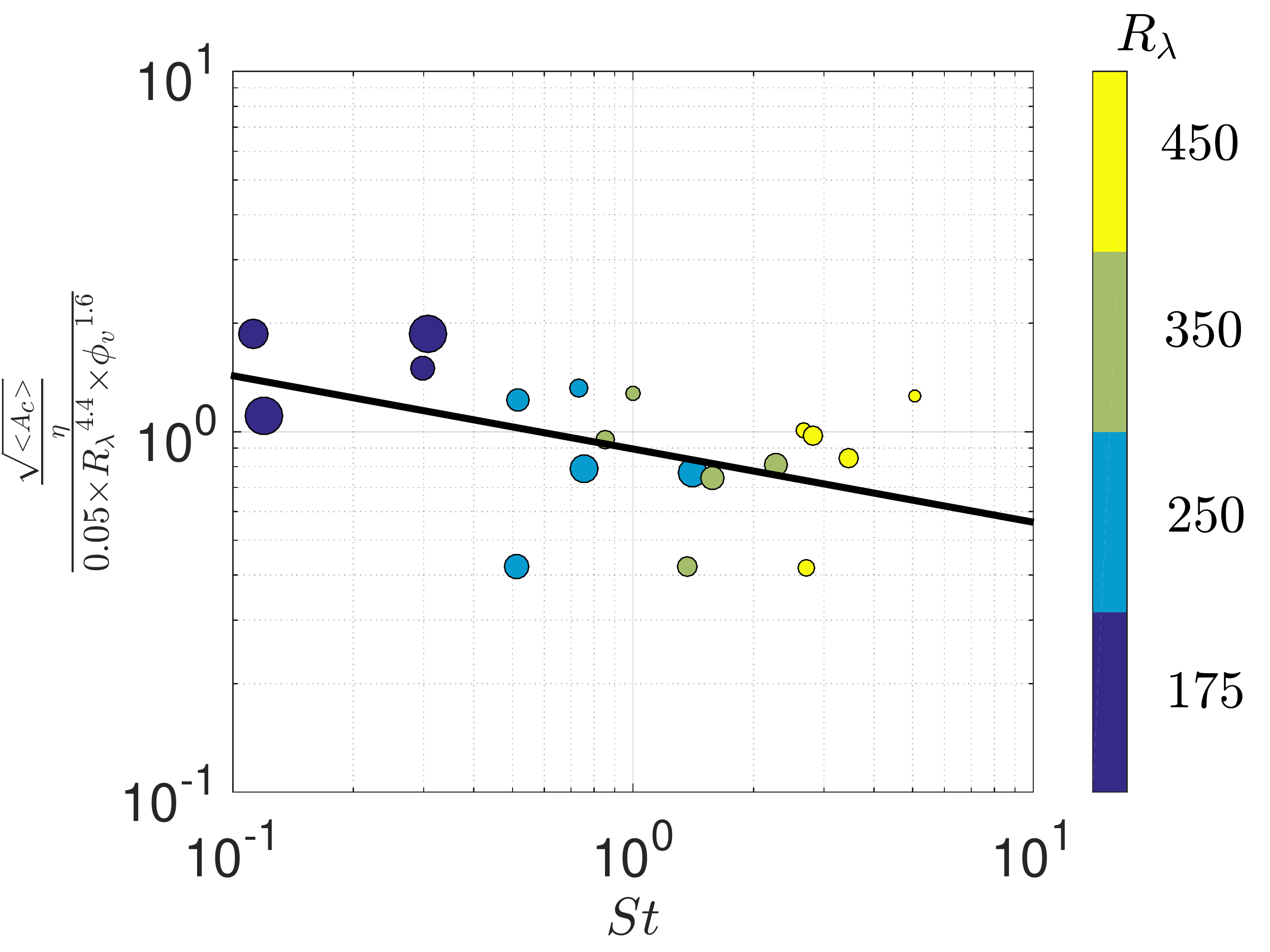}\\
      (a) \hspace{.5\columnwidth} (b)
      \caption{a). Scalings of $\frac{\sqrt{\left<A_c\right>}}{\eta}$ vs $\phi_v$ and $R_\lambda$  b). Reducing the original function of $\frac{\sqrt{\left<A_c\right>}}{\eta}$ shows the weak dependence on $St$}
      \label{fig:scalings_clusterSize}
\end{figure}

Fig.~\ref{fig:clusters_area_voids} shows that the characteristic cluster size vary with the Reynolds number. Fig.~\ref{fig:clusters_StRePhi} quantifies the dependency of the $\frac{\sqrt{\left<A_c\right>}}{\eta}$ with $St$, $Re_{\lambda}$ and $\phi_v$. At first sight, these plots seem to suggest that cluster size increases with increasing Stokes and Reynolds number and decreases with increasing volume fraction. However, as for the previous discussion on $\sigma_{\cal V}$, these trends are quite complex. Fig.~\ref{fig:clusters_StRePhi}a shows that the increase of $\frac{\sqrt{\left<A_c\right>}}{\eta}$ with $St$ is very much connected to that in $Re_\lambda$ (whose value is encoded in the colors of the symbols). Similarly, figs.~\ref{fig:clusters_StRePhi}b\&c also point towards a direct connection between trends of $\frac{\sqrt{\left<A_c\right>}}{\eta}$ on $Re_\lambda$ and $\phi_v$. To obtain better insight into the specific sensitivity to each controlling non-dimensional parameter, power law fits are computed, in the form:
\begin{equation}
	\frac{\sqrt{\left<A_c\right>}}{\eta}=K' St^{\alpha '} Re_\lambda^{\beta '} \phi_v^{\gamma '}
\end{equation}
First, the joint dependencies on $Re_\lambda$ and $\phi_v$, shown in fig.~\ref{fig:scalings_clusterSize}a, are computed.The best fit is obtained for $\beta'=4.4\pm 1.3$ and $\gamma'=1.6\pm0.7$. The dependency of cluster size with volume fraction therefore appears to be marginal compared to the Reynolds number dependency. The remaining dependency on $St$ is then probed by fitting the normalized quantity $\frac{\sqrt{\left< A_c \right>}/\eta}{Re_\lambda^{4.4}\phi_v^{1.6}}$, shown in fig.~\ref{fig:scalings_clusterSize}b. The Stokes number dependency of the cluster size, $\alpha'=-0.2 \pm 0.25$, is relatively weak. Overall, the cluster size dependency on $(St,Re_\lambda,\phi_v)$ can be approximately quantified by the empirical expression:
\begin{equation}
    \frac{\sqrt{\left<A_c\right>}}{\eta}=0.05 \cdot St^{-0.2} Re_\lambda^{4.4} \phi_v^{1.6},
\end{equation}
which shows the dominant role of the Reynolds number, a super-linear dependency on volume fraction and a negligible dependency on Stokes number. This suggests that the cluster size is primarily controlled by the carrier flow turbulence rather than by the disperse phase properties.

Similar trends are also obtained for the size of voids, with sensitivities to $Re_\lambda$ and to $\phi_v$  similar to those obtained for the average cluster dimension. The Stokes number dependency is also weak. Since the spatial extension of the void regions is about ten time larger than that of clusters, this ratio carries into the prefactors in equations (8) and (9).
\begin{equation}
	\frac{\sqrt{\left<A_v\right>}}{\eta}=0.45 \cdot St^{-0.09 \pm 0.1} Re_\lambda^{3.6 \pm 1} \phi_v^{1.3 \pm 0.55 },
\end{equation}

\begin{figure}[t]
	\centering
      \includegraphics[width=.49\columnwidth]{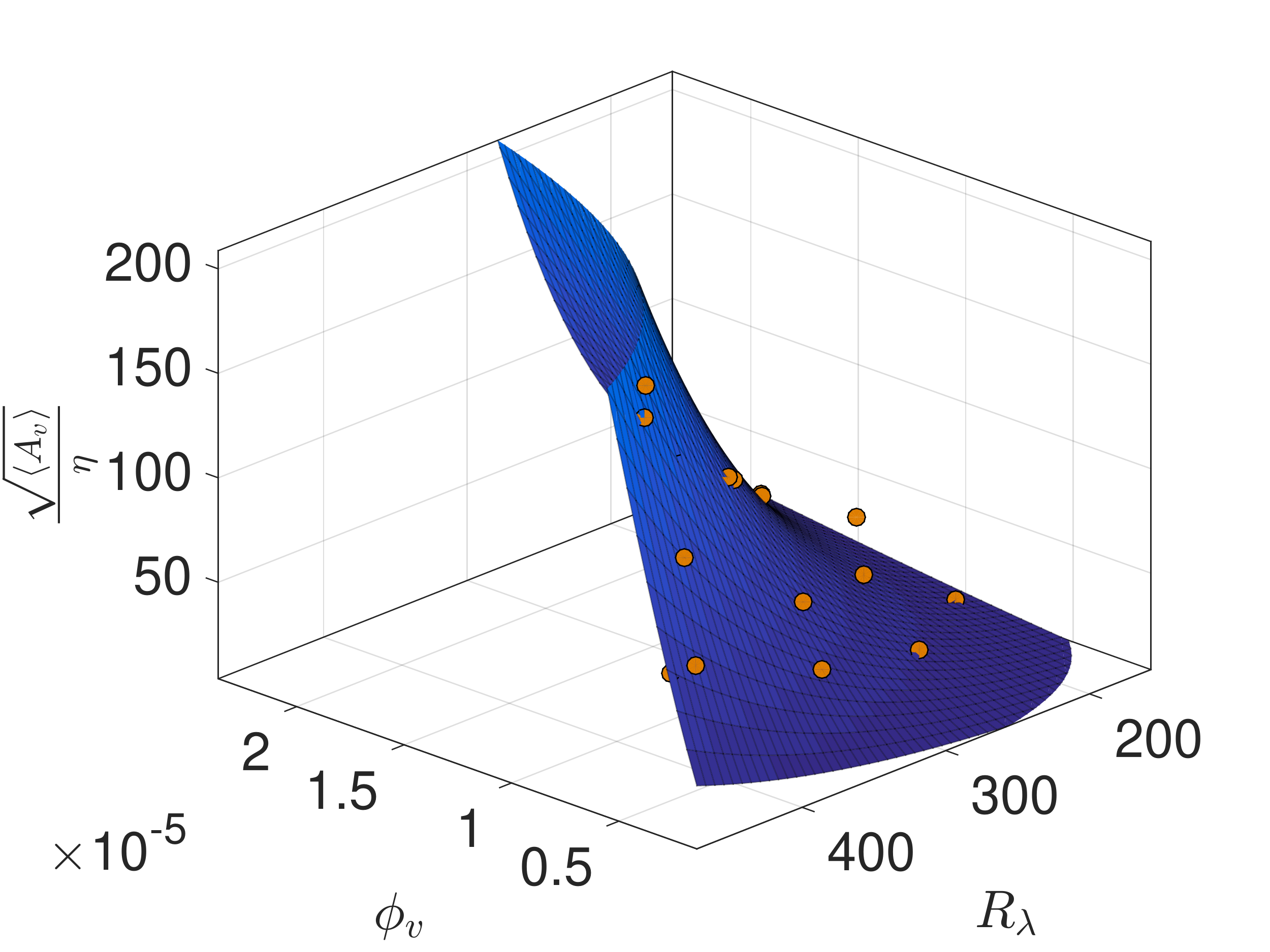}
      \includegraphics[width=.49\columnwidth]{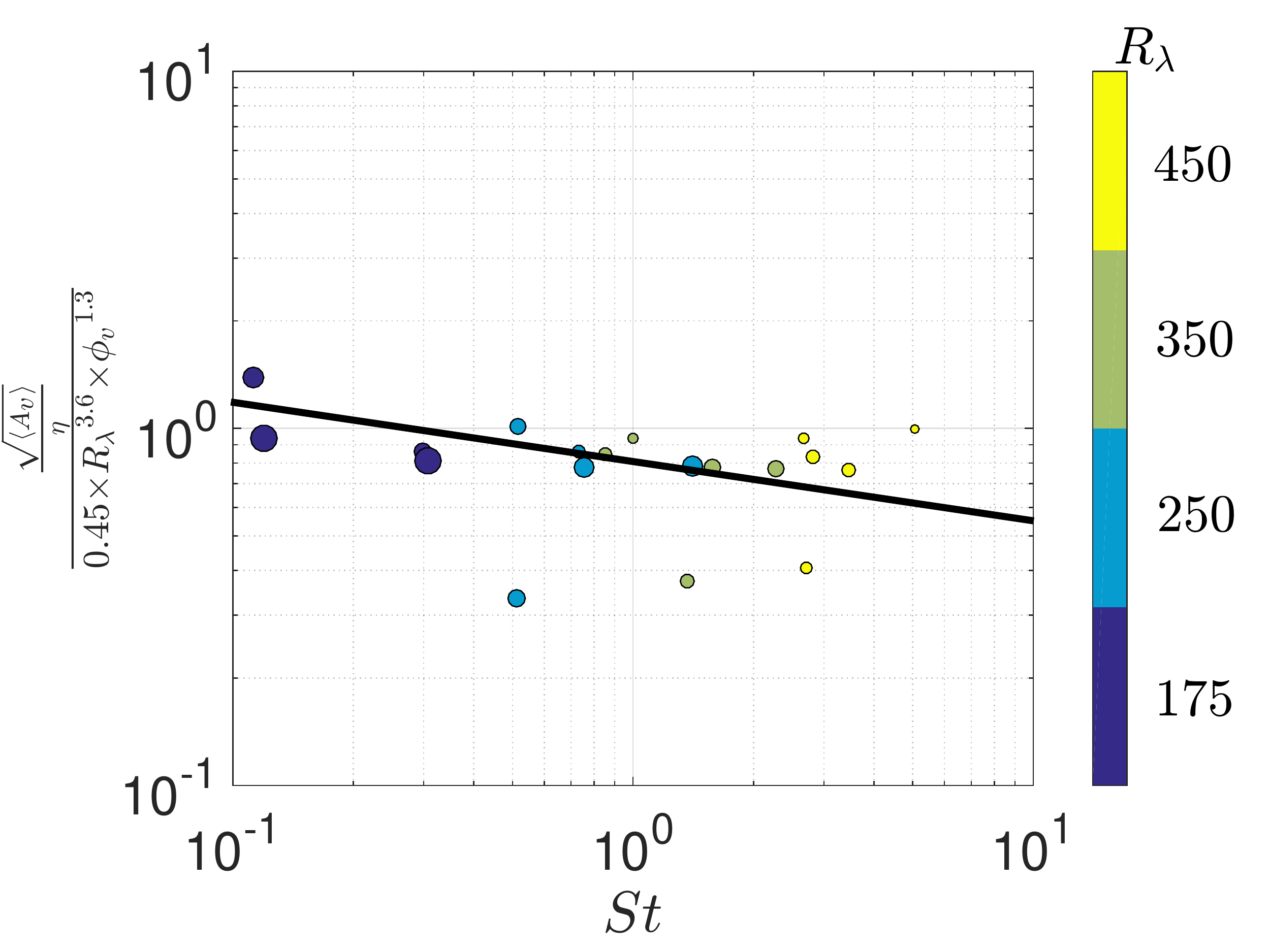}\\
      (a) \hspace{.5\columnwidth} (b)
      \caption{a). Scalings of $\frac{\sqrt{\left<A_v\right>}}{\eta}$ vs $\phi_v$ and $R_\lambda$  b). Reducing the original function of $\frac{\sqrt{\left<A_v\right>}}{\eta}$ shows the weak dependence on $St$ }
      \label{fig:scalings_clusterSizevoids}
\end{figure} 
\section{Discussion}\label{sec:discussion}

Application of Vorono\"i area statistical analysis to quantifying the geometry of cluster and voids of inertial particles in homogeneous isotropic turbulence has revealed the dependence of the preferential concentration on $St$, $Re_{\lambda}$ and volume fraction $\phi_v$. The standard deviation $\sigma$ of the statistics of Vorono\"i areas around particles, as well as the length scales of clusters and voids, have a strong dependency on Reynolds number, an intermediate dependency on volume fraction and no significant dependence (within experimental error) on the Stokes number. 

This strong dependency of clustering on the $Re_{\lambda}$ number reveals the dominant role of carrier flow turbulence in the clustering process, consistent with the assumption that the turbulent structures are the ones responsible for the formation of clusters. 

The dependency of clustering on the particle volume fraction $\phi_v$ is   reminiscent of collective effects due to particle interactions, and is in agreement with previous observations of such collective effects~\cite{bib:aliseda2002_JFM,bib:monchaux2010_PoF}.

A very weak, almost inexistent, dependency of cluster geometry on the Stokes number, based on the maximum probability diameter in the polydisperse particle distributions used in these experiments, has been found. It has been consistently reported from DNS of mono disperse particle-laden flows~\cite{bib:wang1993_JFM,bib:fessler1994_PoF,bib:bec2006_JFM,bib:bec2007_PRL} that clustering is maximum for Stokes number of order unity, invoking a better resonance between particles response time and small turbulent eddies. Most metrics used to characterize the level of clustering are based on small scale quantities, for instance the correlation dimension that measures the increase of probability of finding two particles at vanishing distance compared to a random distribution. Such metrics are only relevant to quantify small scale clustering at sub-dissipative scales, which has been shown to be driven by Reynolds  number and to be essentially independent of Stokes number~\cite{bib:bec2007_PRL}. This analysis is very different to that used in experiments with metrics that focus on inertial scales (most accessible Vorono\"i cells in experiments, such as the one shown in figure~\ref{fig:voroExp}, have dimensions within the inertial range of scales). In line of previous numerical studies~\cite{bib:boffetta2004_PoF,bib:bec2007_PRL,bib:goto2006_PoF}, our experimental results point towards clustering of inertial particles being not only a small scale phenomenon, but one that occurs at all scales of turbulence. This is revealed, for instance, by the algebraic decay of the PDF of cluster areas, and by the fact that average cluster dimensions, up to 100$\eta$, can be found for experiments at the highest Reynolds numbers. The importance of multi-scale clustering has also been recently emphasized by ~\cite{bib:coleman2009_PoF}, who showed that the usual centrifugation mechanism~\cite{bib:maxey1987_JFM}, which is by essence a small scale preferential clustering mechanism based on the negative effective compressibility of high-strain/low vorticity regions of the carrier turbulence, is not the primary mechanism for preferential concentration of particles in turbulence when the Stokes number exceeds unity. Their numerical study shows that for particles with Stokes number larger than unity, clustering is primarily driven by the ``sweep-stick'' mechanism~\cite{bib:coleman2009_PoF} by which particles tend to preferentially sample the zero-acceleration points of the carrier flow. It is important to note that, contrary to the centrifugation mechanism which is indeed clustering mechanism, the ``sweep-stick'' is a preferential sampling mechanism, and clustering only emerges as a consequence of the low-acceleration points in a turbulent flow organizing in multi-scale clusters~\cite{bib:goto2006_PoF}. In this framework, clustering properties are driven by turbulence characteristics across scales, while particle properties only influence the ability of particles to preferentially stick to the aforementioned zero-acceleration points. The main constraint for particles to efficiently stick to zero-acceleration points is that their viscous relaxation time $\tau_p$ be small compared to the life-time of those zero-acceleration points. These points are known from numerical simulations to be very persistent~\cite{bib:goto2006_PoF}, and this can be related to the experimental finding that the correlation time of the acceleration magnitude of tracer particles is of the order of the integral time-scale $T_{int}$ of the carrier turbulence~\cite{bib:mordant2002_PRL}. As a consequence, as long as $\tau_p\ll T_{int}$, no significant dependency of clustering by the ``sweep-stick'' mechanism on the Stokes number is expected. A significant decrease of the efficiency of the mechanism will only occur for particles with response times approaching the integral time scale of the flow. In our experiment, $T_{int}$ is at least of the order of $100$~ms or more. For water droplets, such high response times, would require particles with diameter of the order of 100~$\mu$m or more. Interestingly, the ``sweep-stick'' scenario also suggests that the impact of Stokes number should be more visible at lower Reynolds number as the condition $St \ll T_{int}/\tau_\eta$ becomes more stringent for lower Reynolds numbers. This may explain why, in low Reynolds number simulations~\cite{bib:wang1993_JFM}, where $R_\lambda\approx 30$ and experiments~\cite{bib:monchaux2010_PoF}, a decrease of clustering was indeed observed when Stokes number exceeds unity. 

Finally, we also point out that the polydispersity of our droplet distribution would also be very likely to smear out possible weak Stokes number dependencies, in particular for the experiments at the lowest Reynolds numbers for which some dependency may still have been expected.

\section{Conclusions}\label{sec:conclusions}

Overall, our findings of a dominant role of the Reynolds number compared to the Stokes number is  consistent with a leading multi-scale clustering process driven by a preferential sampling mechanism, such as '' sweep-stick'', in agreement with previous experimental results~\cite{bib:obligado2014_JoT}. In a broader framework, this finding also supports the necessity to distinguish small scale mechanisms of clustering and multi-scale mechanisms~\cite{bib:gustavsson2016_AdvPhys}.

The sensitivity of clustering to the volume fraction identified here is clearly beyond any measurement uncertainty. If, as discussed above, the sweep-stick mechanism is driving the cluster formation, any volume fraction influence is not captured in that picture. A possible scenario could rely on collective effects which are known to lead to denser regions sinking in the mixture with an enhanced settling velocity. Such denser regions could thus collect extra particles during their motion relative to the fluid, and therefore built up clusters of higher concentration and of larger size. Such a process would be clearly favored at higher volume fractions. In this scenario, the sweep-stick mechanism will act as the trigger of cluster formation, with subsequent growth driven by the collective dynamics. Another alternative view is that the presence of clusters modifies the local turbulent structure and favor the multiplication of sticking points in the flow (note that at the largest concentrations in clusters, the mass loading exceeds 0.1 and can even become close to unity): more particles could then either \emph{activate} more zero acceleration points or help bring new particles in the sticking region. These scenarios,  hypothetical as they are, may serve for planing new experiments to help understand how collective effects become efficient in clustering. Clearly, an investigation of the effect of disperse phase volume fraction on the micro scale mechanism for accumulation of particles would be worth undertaking.




We finish by emphasizing that due to intertwining of all three control parameters $St$, $Re_{\lambda}$ and $\phi_v$, the separation of their influence on clustering is an extremely difficult task. More experiments are being conducted to extend quantitative understanding to a broader range of parameter values, in particular regarding the role of volume fraction and collective effects.

The investigation of clustering in regards to its effect on settling of inertial particles is another important aspect that can be studied via conditioned joint statistics of settling velocity and Vorono\"i analysis. This study should ideally provide the dependency of the settling velocity of inertial particles on turbulence fluctuations and a final expression for the connection between settling and clustering. 
    
\section*{Acknowledgments}
This work has been partially supported by the LabEx Tec 21 (Investissements d'Avenir - grant agreement ANR-11-LABX-0030) and by the ANR project TEC2 (grant agreement ANR-12-BS09- 011-03).



\clearpage
\medskip
\bibliographystyle{unsrt}
\bibliography{main}

\begin{thebibliography}{10}

\bibitem{bib:maxey1987_JFM}
M~Maxey.
\newblock {The Gravitational Settling Of Aerosol-Particles In Homogeneous
  Turbulence And Random Flow-Fields}.
\newblock {\em Journal of Fluid Mechanics}, 174:441--465, 1987.

\bibitem{bib:goto2008_PRL}
Susumu Goto and J.~Vassilicos.
\newblock {Sweep-Stick Mechanism of Heavy Particle Clustering in Fluid
  Turbulence}.
\newblock {\em Physical Review Letters}, 100(5):54503, feb 2008.

\bibitem{bib:coleman2009_PoF}
S~W Coleman and J~C Vassilicos.
\newblock {A unified sweep-stick mechanism to explain particle clustering in
  two- and three-dimensional homogeneous, isotropic turbulence}.
\newblock {\em Physics of Fluids}, 21(11):113301, nov 2009.

\bibitem{bib:qureshi2007_PRL}
Nauman~M. Qureshi, Mickael Bourgoin, Christophe Baudet, Alain Cartellier, and
  Yves Gagne.
\newblock {Turbulent transport of material particles: An experimental study of
  finite size effects}.
\newblock {\em Physical Review Letters}, 99(18), 2007.

\bibitem{bib:qureshi2008_EPJB}
N.~M. Qureshi, U.~Arrieta, C.~Baudet, A.~Cartellier, Y.~Gagne, and M.~Bourgoin.
\newblock {Acceleration statistics of inertial particles in turbulent flow}.
\newblock {\em European Physical Journal B}, 66(4):531--536, 2008.

\bibitem{bib:xu2008_PhysicaD}
Haitao Xu and Eberhard Bodenschatz.
\newblock {Motion of inertial particles with size larger than Kolmogorov scale
  in turbulent flows}.
\newblock {\em Physica D}, 237(14-17):2095--2100, 2008.

\bibitem{bib:lucci2011_PoF}
Francesco Lucci, Antonino Ferrante, and Said Elghobashi.
\newblock {Is Stokes number an appropriate indicator for turbulence modulation
  by particles of Taylor-length-scale size?}
\newblock {\em Physics of Fluids}, 23(2):25101, 2011.

\bibitem{bib:homann2010_JFM}
Holger Homann and Jeremie Bec.
\newblock {Finite-size effects in the dynamics of neutrally buoyant particles
  in turbulent flow}.
\newblock {\em JOURNAL OF FLUID MECHANICS}, 651:81--91, 2010.

\bibitem{bib:fiabane2012_PRE}
L~Fiabane, R~Zimmermann, R~Volk, J-F Pinton, and M~Bourgoin.
\newblock {Clustering of finite-size particles in turbulence}.
\newblock {\em Physical Review E}, 86(3):35301, 2012.

\bibitem{bib:doychev2010_ICFM}
T.~Doychev and M.~Uhlmann.
\newblock {A numerical study of finite size particles in homogeneous turbulent
  flow}.
\newblock In S.~Balachandar and J.~Sinclair Curtis, editors, {\em ICMF 2010},
  Proc. 7th Int. Conf. Multiphase Flow, Tampa, USA, 2010. CDROM.

\bibitem{bib:wang1993_JFM}
L~P Wang and M~R Maxey.
\newblock {Settling velocity and concentration distribution of heavy particles
  in homogeneous isotropic turbulence}.
\newblock {\em Journal of Fluid Mechanics}, 256:27--68, 1993.

\bibitem{bib:yang1998_JFM}
C.~Y. Yang and U.~Lei.
\newblock {The role of the turbulent scales in the settling velocity of heavy
  particles in homogeneous isotropic turbulence}.
\newblock {\em Journal of Fluid Mechanics}, 371:179--205, sep 1998.

\bibitem{bib:aliseda2002_JFM}
A~Aliseda, A~Cartellier, F~Hainaux, and J~C Lasheras.
\newblock {Effect of preferential concentration on the settling velocity of
  heavy particles in homogeneous isotropic turbulence}.
\newblock {\em Journal of Fluid Mechanics}, 468:77--105, 2002.

\bibitem{bib:csanady1963_JAtmSc}
G.~T. Csanady.
\newblock {Turbulent Diffusion of Heavy Particles in the Atmosphere}.
\newblock {\em Journal of the Atmospheric Sciences}, 20:201--208, 1963.

\bibitem{bib:Wells83}
M.R. Wells and D.E. Stock.
\newblock {The effect of crossing trajectories on the dispersion of particles
  in a turbulent flow}.
\newblock {\em Journal of Fluid Mechanics}, 136:31--62, 1983.

\bibitem{bib:maxey1983}
Martin~R Maxey and James~J Riley.
\newblock {Equation of motion for a small rigid sphere in a nonuniform flow}.
\newblock {\em Physics of Fluids}, 26(4):883--889, 1983.

\bibitem{bib:obligado2014_JoT}
Martin Obligado, Tomas Teitelbaum, Alain Cartellier, Pablo~D. Mininni, and
  Mickael Bourgoin.
\newblock {Preferential Concentration of Heavy Particles in Turbulence}.
\newblock {\em Journal of Turbulence}, 15(5):293--310, 2014.

\bibitem{bib:monchaux2010_PoF}
R~Monchaux, M~Bourgoin, and A~Cartellier.
\newblock {Preferential concentration of heavy particles: A Voronoi analysis}.
\newblock {\em Physics of Fluids}, 22(10):103304, 2010.

\bibitem{bib:monchaux2012_IJMF}
Romain Monchaux, Mickael Bourgoin, and Alain Cartellier.
\newblock {Analyzing preferential concentration and clustering of inertial
  particles in turbulence}.
\newblock {\em International Journal of Multiphase Flow}, 40:1--18, 2012.

\bibitem{bib:makita1991}
H~Makita and K~Sassa.
\newblock {Active turbulence generation in a laboratory wind tunnel}.
\newblock In {\em Advances in Turbulence 3}, pages 497--505. Springer, 1991.

\bibitem{bib:mydlarski1996_JFM}
Laurent Mydlarski and Zellman Warhaft.
\newblock {On the onset of high-Reynolds-number grid-generated wind tunnel
  turbulence}.
\newblock {\em Journal of Fluid Mechanics}, 320:331--368, 1996.

\bibitem{bib:poorte2002_JFM}
R~Poorte and A~Biesheuvel.
\newblock {Experiments on the motion of gas bubbles in turbulence generated by
  an active grid}.
\newblock {\em Journal of Fluid Mechanics}, 461:127--154, 2002.

\bibitem{bib:ferenc2007_PhysA}
Jarai-Szabo Ferenc and Zoltan Neda.
\newblock {On the size distribution of Poisson Voronoi cells}.
\newblock {\em Physica A-Statistical Mechanics and Its Applications},
  385(2):518--526, 2007.

\bibitem{bib:tagawa2012_JFM}
Yoshiyuki Tagawa, Juli{\'{a}}n~Mart{\'{\i}}nez Mercado, Vivek~N. Prakash,
  Enrico Calzavarini, Chao Sun, and Detlef Lohse.
\newblock {Three-dimensional Lagrangian Vorono{\"{\i}} analysis for clustering
  of particles and bubbles in turbulence}.
\newblock {\em Journal of Fluid Mechanics}, 693:201--215, jan 2012.

\bibitem{bib:uhlmann2014_JFM}
Markus Uhlmann and Todor Doychev.
\newblock {Sedimentation of a dilute suspension of rigid spheres at
  intermediate Galileo numbers: the effect of clustering upon the particle
  motion}.
\newblock {\em Journal of Fluid Mechanics}, 752:310--348, jul 2014.

\bibitem{bib:obligado2015}
Mart{\'{\i}}n Obligado, Nathana{\"{e}}l Machicoane, Agathe Chouippe, Romain
  Volk, Markus Uhlmann, and Micka{\"{e}}l Bourgoin.
\newblock {Path instability on a sphere towed at constant speed}.
\newblock {\em Journal of Fluids and Structures}, 58:99--108, oct 2015.

\bibitem{bib:obligado2015_EPL}
Mart{\'{\i}}n Obligado, Alain Cartellier, and Micka{\"{e}}l Bourgoin.
\newblock {Experimental detection of superclusters of water droplets in
  homogeneous isotropic turbulence}.
\newblock {\em EPL (Europhysics Letters)}, 112(5):54004, 2015.

\bibitem{Bateson12}
C.P. Bateson and A.~Aliseda.
\newblock Wind tunnel measurements of the preferential concentration of
  inertial droplets in homogeneous isotropic turbulence.
\newblock {\em expif}, 52:1373--1387, 2012.

\bibitem{bib:boffetta2004_PoF}
G~Boffetta, F~{De Lillo}, and A~Gamba.
\newblock {Large scale inhomogeneity of inertial particles in turbulent flows}.
\newblock {\em PHYSICS OF FLUIDS}, 16(4):L20--L23, 2004.

\bibitem{bib:goto2006_PoF}
Susumu Goto and J~C Vassilicos.
\newblock {Self-similar clustering of inertial particles and zero-acceleration
  points in fully developed two-dimensional turbulence}.
\newblock {\em Physics of Fluids}, 18(11):115103, nov 2006.

\bibitem{bib:fessler1994_PoF}
Jr~Fessler, Jd~Kulick, and Jk~Eaton.
\newblock {Preferential Concentration Of Heavy-Particles In A Turbulent Channel
  Flow}.
\newblock {\em Physics of Fluids}, 6(11):3742--3749, 1994.

\bibitem{bib:bec2006_JFM}
Jeremie Bec, Luca Biferale, Guido Boffetta, Antonio Celani, Massimo Cencini,
  Alessandra Lanotte, S~Musacchio, and Federico Toschi.
\newblock {Acceleration statistics of heavy particles in turbulence}.
\newblock {\em Journal of Fluid Mechanics}, 550:349--358, mar 2006.

\bibitem{bib:bec2007_PRL}
Jeremie Bec, Luca Biferale, Massimo Cencini, Alessandra Lanotte, Stefano
  Musacchio, and Federico Toschi.
\newblock {Heavy particle concentration in turbulence at dissipative and
  inertial scales}.
\newblock {\em Physical Review Letters}, 98(8):84502, 2007.

\bibitem{bib:mordant2002_PRL}
N~Mordant, J~Delour, E~L{\'{e}}veque, A~Arn{\'{e}}odo, and J-F Pinton.
\newblock {Long time correlations in Lagrangian dynamics: a key to
  intermittency in turbulence}.
\newblock {\em Physical review letters}, 89(25):254502, 2002.

\bibitem{bib:gustavsson2016_AdvPhys}
K.~Gustavsson and B.~Mehlig.
\newblock Statistical models for spatial patterns of heavy particles in
  turbulence.
\newblock {\em Advances in Physics}, 65:1--57, 2016.

\end{thebibliography}
\end{document}